\documentclass[a4paper,11pt]{article}
\pdfoutput=1 % to allow compilation in JCAP
\usepackage{jcappub}
\usepackage{amsmath,amssymb}
\usepackage{bm}
\usepackage{xcolor}
\usepackage{graphicx}
\usepackage{enumitem}
\usepackage{geometry}
\usepackage{xspace}
\usepackage{pdflscape}
\usepackage{placeins}
\usepackage{epstopdf}
\usepackage{comment}
\makeatletter
\gdef\@fpheader{}
\g@addto@macro\bfseries{\boldmath}
\makeatother

\let\oldsqrt\sqrt
\def\sqrt{\mathpalette\DHLhksqrt}
\def\DHLhksqrt#1#2{%
\setbox0=\hbox{$#1\oldsqrt{#2\,}$}\dimen0=\ht0
\advance\dimen0-0.2\ht0
\setbox2=\hbox{\vrule height\ht0 depth -\dimen0}%
{\box0\lower0.4pt\box2}}

\newcommand{\dd}{\mathrm{d}}
\newcommand{\sss}[1]{{\scriptscriptstyle{#1}}}
\newcommand{\boldmathsymbol}[1]{{\ensuremath{\boldsymbol{#1}}}}

\newcommand{\uPl}{\mathrm{Pl}}

\newcommand{\usssPl}{\sss{\uPl}}

\newcommand{\calH}{\mathcal{H}}

\newcommand{\g}{\mathrm{g}}

\newcommand{\Mp}{M_\usssPl}

\newcommand{\beq}{\begin{equation}}
\newcommand{\eeq}{\end{equation}}
\newcommand{\bea}{\begin{equation}\begin{aligned}}
\newcommand{\eea}{\end{aligned}\end{equation}}

\newlength{\wsingfig}
\newlength{\wdblefig}
\newlength{\wquadfig}
\newlength{\wtriplefig}
\newcommand{\Eq}[1]{Eq.~(\ref{#1})}

\newcommand{\Fig}[1]{Fig.~{\ref{#1}}}

\newcommand{\Sec}[1]{Sec.~\ref{#1}}

\def\br{\bar\rho}
\def\bp{\bar p}

\def\doi{http://doi.org}

 \def\e{\mathrm{e}}

\date{today}

\title{Primordial black hole induced gravitational waves in $f(R)$ gravity}

\author[a,b,c]{Theodoros~Papanikolaou}
\author[d,b,c]{Salvatore~Capozziello}

\affiliation[a]{Laboratory of Theoretical and Computational Physics, Department of Physics, University of Patras, 26504 Patras, Greece}

\affiliation[b]{Scuola Superiore Meridionale, Via Mezzocannone 4, I-80134 Napoli, Italy}

\affiliation[c]{Istituto Nazionale di Fisica Nucleare (INFN), Sezione di Napoli, Via Cinthia 21, I-80126 Napoli, Italy}

\affiliation[d]{Dipartimento di Fisica ``E. Pancini", Università di Napoli ``Federico II", Complesso Universitario di Monte S. Angelo Via Cinthia, Ed. N I-80126 Napoli, Italy}

\emailAdd{papaniko@upatras.gr}
\emailAdd{capozziello@na.infn.it}

\abstract{Ultra-light primordial black holes (PBHs) with masses $M_\mathrm{PBH}<5\times 10^{8}\mathrm{g}$ can trigger an early matter-dominated (eMD) era before Big Bang Nucleosynthesis (BBN) and reheat the Universe through their evaporation. Notably, the initial isocurvature in nature PBH energy density fluctuations can induce abundantly gravitational waves (GWs) due to second-order gravitational effects. In this work, we study this induced GW signal within the context of $f(R)$ gravity  investigating the effect of $f(R)$ gravity on the behaviour of scalar perturbations during the PBH-driven eMD era as well as on the source of the second-order induced tensor modes. In particular, we focus on two minimal $f(R)$ models, that is  $R^2$ and $R^{1+\epsilon}$, with $\epsilon\ll 1$, gravity theories as illustrative examples, finding at the end that $R^2$ gravity presents very small deviations from general relativity (GR) at the level of both the scalar and tensor perturbations. However, $R^{1+\epsilon}$ gravity features a different behaviour, exhibiting in particular an exponential growth of scalar perturbations during a matter-dominated era. This unique feature leads ultimately to an enhanced induced GW signal even for very small initial PBH abundances, being characterized by a linear frequency scaling on large infrared scales away from the peak frequency, namely $\Omega_\mathrm{GW}(k)\propto f$ for $f\ll f_\mathrm{peak}$ and a very peaky behaviour with a $f^{35/2}$ frequency scaling close to the non-linear cut-off scale below which perturbation theory breaks down. Interestingly, one finds as well that the induced GW signal can be well within the sensitivity curves of GW detectors, namely that of LISA, ET, BBO and SKA.
}

\begin{document}

\maketitle

%%%%%%%%%%%%%%%%%%%%%%%% Section 1:  Introduction %%%%%%%%%%%%%%%%%%%%%%%%%%%%%%%%%%%%%%%%%%%%%%%%%%%%%%%%%%%%%%%%
\section{Introduction} 
Primordial black holes (PBHs), firstly introduced in the early `70s~\cite{1967SvA....10..602Z, Carr:1974nx,1975ApJ...201....1C,1979A&A....80..104N}, typically form in the very early Universe before the epoch of star formation through various mechanisms associated to the collapse of enhanced primordial cosmological perturbations~\cite{Carr:1974nx,Carr:1975qj,Yokoyama:1995ex,Garcia-Bellido:2017mdw,Musco:2020jjb}, cosmological phase transitions~\cite{Hawking:1982ga,Moss:1994iq,Jung:2021mku}, scalar field instabilities~\cite{Khlopov:1985fch}, collisions of highly energetic particles~\cite{Saini:2017tsz}, topological defects~\cite{Hawking:1987bn,Polnarev:1988dh} and modified/quantum gravity constructions~\cite{Barrow:1996jk,Kawai:2021edk,Papanikolaou:2023crz}. They have recently rekindled the interest of the scientific community since they can address a plethora of modern cosmological conundra~\cite{Carr:2019kxo,Carr:2023tpt}. Notably, PBHs can account for a part or all of the dark matter bulk
\cite{Chapline:1975ojl,Belotsky:2014kca} while, at the same time, they can potentially explain 
the large-scale structure formation through Poisson 
fluctuations \cite{Meszaros:1975ef,Delos:2024poq}. Moreover, they can 
offer us the seeds for the supermassive black holes residing in galactic centers ~\cite{1984MNRAS.206..315C, Bean:2002kx}, as well as account for some of the black-hole merging events detected by the LIGO/VIRGO/KAGRA collaboration~\cite{Sasaki:2016jop,Franciolini:2021tla}. For recent reviews on PBHs see here~\cite{Khlopov:2008qy,Carr:2020gox,Escriva:2022duf}.

Given the huge progress reported in the recent years in the fields of gravitational-wave astronomy, PBHs can be associated with numerous GW signals~\cite{LISACosmologyWorkingGroup:2023njw,Domenech:2025ior}. Most importantly, PBHs have been related to the so-called scalar-induced GW (SIGW) background since the same enhanced scalar perturbations, responsible for PBH formation, can induce as well primordial GWs at second order in cosmological perturbation theory~\cite{Matarrese:1992rp,Matarrese:1993zf,Matarrese:1997ay,Mollerach:2003nq,Saito_2009,Bugaev:2009zh, Nakama_2015, Odintsov:2025oiy, Oikonomou:2024zhs, Oikonomou:2023qfz, Oikonomou:2022irx, Odintsov:2021kup}[See~\cite{Domenech:2021ztg} for a recent 
review on SIGWs]. Furthermore, there have been extensively studied the GW background associated to PBH merging events 
 \cite{Nakamura:1997sm, Ioka:1998nz, 
Eroshenko:2016hmn, Raidal:2017mfl, Zagorac:2019ekv,Hooper:2020evu} as well as the GW background due to the emission of PBH Hawking radiated-gravitons~\cite{Anantua:2008am,Dong:2015yjs,Ireland:2023avg,Ireland:2023zrd,Wang:2025lti}. 

Apart from the aforementioned PBH associated GW signals, it has been also proposed in~\cite{Papanikolaou:2020qtd,Domenech:2020ssp,Papanikolaou:2022chm}, that GWs can  be induced by isocurvature PBH energy density perturbations due to the inhomogeneous distribution of PBHs at distances much larger than the PBH mean separation scale~\cite{Domenech:2023jve}.  Astonishingly, these GWs can be abundantly produced during an eMD era driven by ultralight PBHs with masses $M_\mathrm{PBH}<5\times 10^8\mathrm{g}$, which reheat the Universe through their evaporation with the latter taking place before BBN~\cite{GarciaBellido:1996qt, Hidalgo:2011fj, Zagorac:2019ekv}. Remarkably, these ultra-light PBHs are associated with a very rich phenomenology, among others dark matter production~\cite{Chianese:2019kyl,Samanta:2021mdm}, the probe of primordial non-Gaussianities on very small scales~\cite{Papanikolaou:2024kjb,He:2024luf} as well as of modified Hawking evaporation ~\cite{Balaji:2024hpu,Chianese:2024rsn}, high-scale leptogenesis~\cite{Calabrese:2023key}, the alleviation of the Hubble tension~\cite{Papanikolaou:2023oxq} and the generation of an early inflationary era accounting for the horizon and the flatness cosmological problems~\cite{Dialektopoulos:2025mfz}.

Interestingly, the GW background signal induced by PBH isocurvature energy density perturbations, has been also studied within the context of modified gravity theories~\cite{Papanikolaou:2021uhe,Papanikolaou:2022hkg} which are mainly introduced at  theoretical level so as to improve renormalizability issues of general relativity~\cite{Stelle:1976gc,Addazi:2021xuf} while, at  phenomenological level, in order to account for the two epochs of cosmic acceleration, namely the early-time, inflationary one 
\cite{Nojiri:2010wj,Martin:2013tda}, and/or the late-time, dark-energy one~\cite{Capozziello:2002rd,Bamba:2012cp,CANTATA:2021ktz,Cai:2015emx}. In principle, these theories of gravity converge to standard GR  at a given limit, but in general they are characterized by a richer structure and new 
degrees of freedom leaving a distinctive impact in the Universe evolution. 

In particular, one can discriminate between two big classes of modified gravity theories. On the one hand, one can deal with \textit{Extended Theories of Gravity} constructed by curvature invariants where GR foundations are retained while the Einstein theory is recovered at some limit~\cite{Capozziello:2011et}. On the other hand, if the foundations of gravitational interaction are approached in different ways like in teleparallel  or in non-metric gravity, one deals with \textit{Alternative Gravity Theories}~\cite{Cai:2015emx, CANTATA:2021ktz}. In these alternative paradigms, curvature has not anymore the central role as in GR to dictate dynamics, with the latter being described in terms of torsion and non-metricity invariants. Besides, the metric description of spacetime is substituted by other geometric objects, called tetrads~\cite{Aldrovandi:2013wha, BeltranJimenez:2019esp}. 

In this work, we will focus on $f(R)$ gravity, where the Einstein-Hilbert action is  a function of the Ricci scalar curvature $R$~\cite{Capozziello:2004vh,Sotiriou:2008rp,DeFelice:2010aj}. In particular, with regard to inflation, a particular  $f(R)$ gravity model known as the Starobinsky one, or $R^2$ gravity~\cite{Starobinsky:1980te}, proves to be one of the  best-fitting model of the Cosmic Microwave Background (CMB) cosmological data~\cite{Planck:2018jri} while at the same time there have been substantial work connecting $f(R)$ gravity  with dark energy~\cite{Capozziello:2002rd, Capozziello:2005ku, Capozziello:2007ec, Hu:2007nk}, dark matter \cite{Capozziello:2012ie} and galactic dynamics~\cite{Capozziello:2006ph, Capozziello:2008ny, Martins:2007uf, Napolitano:2012fp, Mendoza:2006hs, Sobouti:2006rd}. Recently, there has been reported as well some progress  regarding the effect of $f(R)$ gravity  on SIGWs, in particular, by studying the $f(R)$-induced modifications at the level of the source and the propagation of the SIGW signal~\cite{Zhou:2024doz,Kugarajh:2025rbt,Lopez:2025gfu}.  Relevant effects can emerge also when electromagnetic fields are present in the $f(R)$ background as discussed in \cite{Capozziello:2025htb}.

Given thus the strong connection of SIGWs and PBHs as described above, we study, in this paper, the GW signal induced by PBH isocurvature within the context of $f(R)$ gravity  by extending a previous work~\cite{Papanikolaou:2021uhe}. In particular, here we account for the full modified source of the SIGWs in $f(R)$ gravity being characterized by an extra geometry-induced anisotropic stress as well as for the $f(R)$ evolution of the scalar perturbations during the PBH-driven eMD era, being associated, as we will see in the following, with a very rich phenomenology. We also account, for the first time, for the effect of the  scalaron massive mode on the first-order GW background.

In order to illustrate the aforementioned $f(R)$ gravity effects on the PBH induced GW signal, we choose to work with representative classes of power-law $f(R)$ gravity. Namely, we work with $R^2$ Starobinsky gravity, leading to a viable inflation scenario~\cite{Starobinsky:1980te} compatible with Planck observations as well with $R^{1+\epsilon}$ gravity, with $\epsilon\ll 1$, naturally accounting for regularization and renormalization issues of GR up to one-loop level~\cite{Birrell:1982ix, Odintsov:2024lid, Acquaviva:2022bju, Dixit:2024grt, Kruglov:2023odm, Buchbinder:1992gdx}. Interestingly, $R^{1+\epsilon}$ gravity can also account for the clustered galactic dark matter problem~\cite{Capozziello:2007ec,Sharma:2020vex} as well as for anisotropic compact objects~\cite{Astashenok:2020qds,Capozziello:2015yza,Pretel:2022plg} while being viewed as a minimal logarithmic deviation from GR, it is capable to alleviate the $H_0$ and the $\sigma_8$ cosmological tensions~\cite{Gomez-Valent:2017idt,SolaPeracaula:2021gxi}.

The paper is organized as follows: In \Sec{sec:f_R_gravity_basics} we review the fundamentals of $f(R)$ gravity while in \Sec{sec:PBH_gas} we introduce a population (``gas") of PBHs randomnly distributed in space giving rise to isocurvature PBH matter density perturbations. Followingly, in \Sec{sec:scalar_perturbations} we study thoroughly the evolution of scalar perturbations experiencing a eMD era driven by PBHs considering two illustrative $f(R)$ gravity models, the $R^2$ and $R^{1+\epsilon}$ theories with $\epsilon\ll 1$, being motivated above. Then, in \Sec{sec:SIGW}, we explore the SIGW signal deriving ultimately its dominant contributions  and accounting also for the effect of the scalaron massive mode. Finally, in \Sec{sec:results}, we present our results while \Sec{sec:conclusions} is devoted to discussion and conclusions.

%%%%%%%%%%%%%%%%%%%%%%%%%%%% SECTION 2 %%%%%%%%%%%%%%%%%%%%%%%%%%%%%%%%%
\section{The basics of \texorpdfstring{$f(R)$}{f(R)} gravity}\label{sec:f_R_gravity_basics}

Let us  consider a theory where the Einstein-Hilbert action is promoted to a function of the Ricci scalar curvature $R$, 
\begin{equation} 
S = \frac{1}{16\pi G} \int d^4x \sqrt{-g} \, f(R) + \int d^4x \sqrt{-g} 
\mathcal{L}_\mathrm{m} \label{fr},
\end{equation}
where $\mathcal{L}_\mathrm{m} $ is the Lagrangian of the matter sector. This is the so-called $f(R)$ gravity.

By varying the action (\ref{fr}) with respect to the metric $g^{\mu \nu}$, one gets the gravitational field equations reading as:
\begin{equation}
 FR_{\mu \nu} - \frac{1}{2} g_{\mu \nu} f + (g_{\mu \nu} \Box - 
\nabla_{\mu}\nabla_{\nu})F = 8\pi G T^\mathrm{m}_{\mu \nu} \label{sf},  
\end{equation}
where $T^\mathrm{m}_{\mu \nu}$ is the total matter stress-energy tensor \cite{Faraoni:2010pgm}.
%where we have set $F \equiv \mathrm{d}f(R)/\mathrm{d}R$. 
An important feature of the richer structure of  $f(R)$ gravity is the existence of an extra massive degree of freedom, the so-called scalaron field~\cite{Starobinsky:1980te}. One can straightforwardly derive its equation of motion by taking the trace of (\ref{sf}), obtaining finally a wave equation for $\phi_{\mathrm{sc}} \equiv \frac{\mathrm{d}f(R)}{\mathrm{d}R}$ which can be written as
\begin{equation}
    \Box \phi_{\mathrm{sc}} =  \frac{\mathrm{d}V_\mathrm{eff}}{\mathrm{d}F}, \label{eq:scalaron_background}
\end{equation}
where $\frac{\mathrm{d}V_\mathrm{eff}}{\mathrm{d}F}\equiv \frac{1}{3}\left[ 2f(R) - F(R)R + 8\pi G \, T^\mathrm{m} \right]$, $F(R)\equiv \frac{\mathrm{d}f(R)}{\mathrm{d}R}$ and $T^\mathrm{m}$ accounts for the trace of the total matter energy-momentum tensor. It is worth noticing the dynamical nature of the  trace equation which, in GR, is simply $R=-8\pi G\, T^\mathrm{m}$. This is the reason of the further degree of freedom of the $f(R)$ gravity theory. One can also define the mass of the scalar degree of freedom through the effective potential $V_\mathrm{eff}$ introduced above, reading as
\begin{equation}
    m_{\mathrm{sc}}^2  \equiv \frac{\mathrm{d}^2 V_\mathrm{eff}}{\mathrm{d}F^2} = \frac{1}{3}\left( \frac{F}{F_{,R}} - R \right), \label{scalmass}
\end{equation}
where $F_{,R} \equiv \mathrm{d}F/\mathrm{d}R = \mathrm{d}^2f/\mathrm{d}R^2$. 

For our purposes, we consider a cosmological framework, working with the flat Friedmann - Lema\^itre - Robertson -Walker (FLRW) background metric $\mathrm{d}s^{2}_\mathrm{b}$ reading as
\begin{equation}
\label{FLRWmetric_background}
\mathrm{d}s^{2}_\mathrm{b}=-\mathrm{d}t^{2}+a^{2}(t)\delta_{ij}\mathrm{d}x^{i}\mathrm{d}x^{j}\,,
\end{equation}
where $a(t)$ is the scale factor. Ultimately, one can 
extract, from \Eq{sf}, the Friedmann equations within $f(R)$ gravity in a form similar to that of GR. Doing so, it is useful to formulate $f(R)$ gravity modifications of the  Friedmann equations in terms of an effective curvature-induced fluid. In particular, one can rewrite the equations (\ref{sf}) as the corresponding ones in classical gravity, with the introduction of the \textit{curvature-induced  energy-momentum tensor} \cite{Arjona:2018jhh, Capozziello:2021wwv}:  
\begin{align}
   T^\mathrm{f(R) \, \mu }_{\nu} &\equiv (1 - F) R^{\mu}_{\nu} + 
\frac{1}{2}\delta^{\mu}_{\nu}( f - R) - (\delta^{\mu}_{\nu} \Box - 
\nabla^{\mu}\nabla_{\nu})F \label{tfr}.
\end{align}
At the end, the $f(R)$ gravity Friedmann equations can be recast as
\begin{align}
\mathcal{H}^{2} &=\frac {8\pi G a^2}{3} \br_\mathrm{tot} \label{FR1}\\
 \mathcal{H} ' &=-\frac{4\pi G a^2}{3}  (\br_\mathrm{tot} + 3\bp_\mathrm{tot}),
 \label{FR2}
\end{align}
where prime $'$ denotes differentiation with respect to the conformal time $\eta$ defined as $\mathrm{d}\eta \equiv a\mathrm{d}t$.
We need to note however that, in this case, in the total energy budget of the Universe one should consider the contribution of the $f(R)$ curvature-induced effective fluid introduced above, whose energy density and pressure are given by \cite{DeFelice:2010aj, Capozziello:2011et}:
\begin{align}
 \br_\mathrm{f(R)} &\equiv -T^{\mathrm{f(R)} \,0 }_{0} = \frac{1}{8\pi G a^2} \Big( 3\mathcal{H}^2- \frac{1}{2} a ^2 f + 3F\mathcal{H}'- 3\mathcal{H} F' \Big) \label{reff}\\
 \bp_\mathrm{f(R)} &\equiv \frac{T^{ \mathrm{f(R)} \,i }_{ i}}{3} = \frac{1}{8\pi G a^2} \Big(- 2\mathcal{H}'- \mathcal{H}^2 + \frac{1}{2} a ^2 f - F\mathcal{H}'- 2F \mathcal{H}^2 + F'' + \mathcal{H} F' \Big). \label{peff}
\end{align}

%%%%%%%%%%%%%%%%%%%%%%%%% SECTION 3 %%%%%%%%%%%%%%%%%%%%%%
\section{The primordial black hole gas}\label{sec:PBH_gas}
\subsection{Early matter dominated eras driven by primordial black holes}
Let us now consider a population (``gas") of PBHs forming during the radiation-dominated (RD) era after inflation  and characterized for simplicity by a PBH mass distribution peaking around a characteristic scale $k_\mathrm{f}$. Neglecting non-Gaussianities, we assume Gaussian curvature perturbations collapsing to form PBHs. One then is inevitably met with a random PBH spatial distribution~\cite{Desjacques:2018wuu, Ali-Haimoud:2018dau, MoradinezhadDizgah:2019wjf}. In order not to affect the production of light elements taking place during BBN, we require as well that PBHs evaporate by BBN time, imposing thus an upper limit on their mass of the order of $10^9\mathrm{g}$~\cite{Kawasaki:1999na,Kawasaki:2000en,Hasegawa:2019jsa,Carr:2020gox}. 

Interestingly, the PBH abundance $\Omega_\mathrm{PBH}$ will evolve as $\Omega_\mathrm{PBH}\propto \rho_\mathrm{PBH}/\rho_\mathrm{r}\propto a^{-3}/a^{-4}\propto a$ since we have a matter constituent (PBHs in our case) evolving in a RD background~\footnote{Such early-matter dominated eras driven by ultra-light black holes or non-relativistic particles have been extensively studied in the literature being associated with a very rich phenomenology~\cite{Khlopov:1980mg,1981SvA....25..406P,1982AZh....59..639P,Polnarev:1985btg,GarciaBellido:1996qt,Hidalgo:2011fj,Suyama:2014vga,Zagorac:2019ekv,Chattopadhyay:2022fwa,Pearce:2025ywc}.}  Finally, if the initial PBH abundance at PBH formation time $\Omega_\mathrm{PBH,f}$ is high enough, PBHs will dominate at a later stage the energy content of the Universe  with the scale factor at PBH domination time reading as $a_\mathrm{d} = a_\mathrm{f}/\Omega_\mathrm{PBH,f}$ and with the scale factor at PBH formation time $a_\mathrm{f}$ being recast as~\cite{Domenech:2020ssp}
\beq\label{eq:scale_factor_PBH_formation}
a_\mathrm{f} = \left(\frac{A\Mp^2}{2\pi\gamma\Omega^2_\mathrm{PBH,f}M^2_\mathrm{PBH}}\right)^{1/6}{\Omega^{(0)}_\mathrm{r}}^{1/4}\sqrt{\frac{H_0M_\mathrm{PBH}}{4\pi\gamma\Mp^2}},
\eeq
where $H_0\simeq 70\mathrm{km/s/Mpc}$ is the Hubble parameter today, $\Omega^{(0)}_\mathrm{r} = 4\times 10^{-5}$ is the present-day abundance of radiation and $A= 3.8\times g_\mathrm{eff}/960$, with $g_\mathrm{eff}\simeq 100$~\cite{Kolb:1990vq} being the effective number of relativistic degrees of freedom present at the epoch of PBH formation. In order for this early matter-dominated (eMD) era driven by light PBHs to take place before PBH evaporation, one needs to impose a lower bound on the PBH abundance at formation reading as~\cite{Papanikolaou:2020qtd}
\beq
\Omega_\mathrm{PBH,f}\geq 10^{-10}C_\mathrm{G}\left(\frac{g_\mathrm{eff}}{108}\right)^{1/2}\left(\frac{10^4\mathrm{g}}{M_\mathrm{PBH}}\right),
\eeq
where $C_\mathrm{G}$ is a parameter depending on the underlying gravity theory within which PBHs form. According to numerical searches in the field, the evaporation time of PBHs in modified gravity, in particular those which are regular in the center, is larger than the one of GR due to the richer structure of the theory, typically one order of magnitude larger or less~\cite{Calza:2024fzo,Calza:2024xdh,Calza:2025mwn} leading to $t^\mathrm{G}_\mathrm{evap}=\frac{t^{\mathrm{GR}}_\mathrm{evap}}{C_\mathrm{G}}$ with $C_\mathrm{G}<1$. For our numerical applications, we choose to be conservative and choose $C_\mathrm{G}=1/5$, leading to non considerable effects on the maximum PBH mass and abundances compatible with BBN.

\subsection{The primordial black hole gravitational potential in general relativity}\label{sec:P_Psi_PBH}
Having introduced above the PBH-dominated Universe, one can view PBHs at distances much larger than the PBH mean separation scale, behaving effectively as a pressureless matter fluid characterized with its own
energy density perturbations $\delta_\mathrm{PBH}$, being of isocurvature nature~\cite{Inman:2019wvr,Papanikolaou:2020qtd}. When  these ultralight PBHs will dominate the energy budget of the Universe as described above, their isocurvature energy density perturbations will convert to curvature perturbations deep in the PBH-driven eMD era~\cite{Kodama:1986fg,Kodama:1986ud}, associated with a gravitational potential $\Psi_\mathrm{PBH}$ \`a la Bardeen~\footnote{Note that at early times when the PBH-driven eMD era occurs, there is no anistotropic stress. Thus, $\Psi_\mathrm{PBH} = \Phi_\mathrm{PBH}$. This is not necessarily the case as we will see later for $f(R)$ gravity due to the emergence of an anisotropic stress of geometric origin.}. This PBH gravitational potential is basically associated with the PBH number density fluctuations. One can easily show by treating PBHs as discrete objects that the two-point correlation function for the PBH density perturbations $\delta_\mathrm{PBH}$ read as
\begin{equation}
    \langle\delta_{\rm PBH}({\bm x}_1)\delta_{\rm PBH}({\bm x}_2)\rangle=\frac{4\pi}{3}\left(\frac{\bar r}{a}\right)^3\delta_{\rm D}(\bm x_1-\bm x_2)\,,
\end{equation}
where $\bm x$ stands for comoving coordinates, and $\bar r$ denotes the PBH mean separation scale, written as $\displaystyle{\bar r = (\tfrac{3M_{\rm PBH}}{4\pi\rho_{\rm PBH}})^{1/3}}$. At scales smaller than $\bar r$, PBHs cannot be described anymore as pressureless dust, something which gives us an ultra-violet (UV) cutoff reading as~\cite{Papanikolaou:2020qtd}
\begin{equation}
    k_{\rm UV}\equiv \frac{a}{\bar r}=\frac{a_{\rm f}}{{\bar r}_{\rm f}}=\gamma^{1/3}\Omega^{1/3}_\mathrm{PBH,f}k_{\rm f},
\end{equation}
where $\Omega_\mathrm{PBH,f}$ is the PBH abundance at formation and $\gamma$ is the fraction of the cosmological horizon collapsing to PBHs. For PBHs forming during the RD era, $\gamma \simeq 0.36$~\cite{Musco:2008hv}.

Going to $k$ space, the reduced PBH matter power spectrum defined as ${\cal P}(k)\equiv k^3|\delta_k|^2/(2\pi^2)$ will be recast as~\cite{Papanikolaou:2020qtd}
\begin{equation}\label{eq:ppoi}
    {\cal P}_{\delta_\mathrm{PBH}}(k)=\frac{2}{3\pi}\left(\frac{k}{k_{\rm UV}}\right)^3,
\end{equation}
with the UV cutoff scale $k_\mathrm{UV}$ being the scale below which, one enters the non-linear regime, where perturbation theory breaks down, namely $ {\cal P}_{\delta_\mathrm{PBH}}(k>k_\mathrm{UV})>\mathcal{O}(1)$, as it can be checked by \Eq{eq:ppoi}. We will consider thus followingly scales larger than the mean PBH separation scale, i.e. $k<k_\mathrm{UV}$, where the PBH gas can be effectively treated as a pressureless dust fluid and where perturbation theory is valid.

Finally, accounting for the initial isocurvature nature of $\delta_\mathrm{PBH}$, the linear growth of $\delta_\mathrm{PBH}\propto a$ during the PBH-driven matter era and the behaviour of $\Phi$ and the comoving curvature perturbation on super and sub-horizon scales, one can show after a long calculation [See \Sec{sec:Psi_Phi_PBH_f_R_gravity} for the case of $f(R)$ gravity] that the power spectrum of the PBH gravitational potential at the end of the PBH-dominated era will read as~\cite{Domenech:2020ssp}:
\beq\label{eq:P_Phi_PBH_full}
\mathcal{P}_{\Psi_{\mathrm{PBH}}}(k)=S^2_{\Psi_\mathrm{PBH}}(k)\left(5+\frac{8}{9}\frac{k^2}{{k}^2_\mathrm{d}}\right)^{-2}\mathcal{P}_{\delta_\mathrm{PBH}}(k)\,,
\eeq
where $S_{\Psi_\mathrm{PBH}}(k) \simeq \left(\sqrt{\frac{2}{3}}\frac{k}{k_\mathrm{evap}}\right)^{-1/3}$ is a factor suppressing scales with time variation larger than the PBH evaporation time due to the non-zero pressure of the radiation fluid~\cite{Inomata:2020lmk}. $k_\mathrm{d}$ and $k_\mathrm{evap}$ are the scales crossing the cosmological horizon at the beginning and at the end of the PBH-dominated era respectively.

Notably, the PBH induced metric perturbations $\Psi_\mathrm{PBH}$ can induce an abundant production of GWs generated due to second order gravitational interactions in cosmological perturbation theory being potentially detectable in future GW observatories such as LISA, ET and BBO~\cite{Papanikolaou:2020qtd, Domenech:2020ssp,Papanikolaou:2022chm}.

\section{Scalar perturbations}\label{sec:scalar_perturbations}
Let us then study in this Section the behaviour of the PBH induced scalar metric perturbations within the context of $f(R)$ gravity. Starting with the metric, one can write generically its perturbed form considering only the first order scalar perturbations  $\Psi^{(1)}$ and $\Phi^{(1)}$ as
\begin{equation}\label{eq:ds}
\begin{aligned}
	\mathrm{d} s^2  =a^2(\eta)\left[-\left(1+2 \Phi^{(1)}\right) \mathrm{d} \eta^2 +\left(1-2 \Psi^{(1)}\right) \delta_{i j} \mathrm{d} x^i \mathrm{~d} x^j\right],
\end{aligned}
\end{equation}
by keeping both the $\Phi^{(1)}$ and $\Psi^{(1)}$ Bardeen gravitational potentials and where the indices $(1)$ denote first order perturbed quantities respectively. We should note here that generically $\Phi^{(1)}$ and $\Psi^{(1)}$ are not necessarily equal \cite{Capozziello:2007ms}.

One then can show that for a generic $f(R)$ gravity model the dynamical evolution of the first-order Bardeen gravitational potentials $\Phi^{(1)}$ and $\Psi^{(1)}$ are governed by the following equations~\cite{Tsujikawa:2007gd}:
\beq
\begin{split}\label{eq:Phi_Psi}
2\mathcal{H}&\left[3(c^2_\mathrm{s,tot}- w_\mathrm{tot})\mathcal{H}\Phi^{(1)}  + \Phi^{(1)\prime}+\left(2+3c^2_\mathrm{s,tot}\right)\Psi^{(1)\prime}\right] + 2\Psi^{(1)\prime\prime} \\ & +\Delta\Phi^{(1)} -\Delta\Psi^{(1)} - 2c^2_\mathrm{s,tot} \Delta\Psi^{(1)} = \frac{a^2\delta P_\mathrm{iso}}{2\Mp^2}%\bar{\rho}_\mathrm{m}c^2_\mathrm{s}S,
\end{split}
\eeq
\beq\label{eq:Psi_minus_Phi}
\Psi^{(1)} - \Phi^{(1)} = \frac{F^{(1)}}{F^{(0)}},
\eeq
where $w_\mathrm{tot}\equiv \frac{\bar{p}_\mathrm{tot}}{\bar{\rho}_\mathrm{tot}}$ and $c^2_\mathrm{s,tot}\equiv \frac{p^{(1)}_\mathrm{tot}}{\rho^{(1)}_\mathrm{tot}}$ with $\bar{\rho}_\mathrm{tot} \equiv \bar{\rho}_\mathrm{m} + \bar{\rho}_\mathrm{f(R)}$, $\rho^{(1)}_\mathrm{tot} \equiv \rho^{(1)}_\mathrm{m} + \rho^{(1)}_\mathrm{f(R)}$ and $\bar{p}_\mathrm{tot} \equiv \bar{p}_\mathrm{m} + \bar{p}_\mathrm{f(R)}$ and $p^{(1)}_\mathrm{tot} \equiv p^{(1)}_\mathrm{m} + p^{(1)}_\mathrm{f(R)}$. $\delta P_\mathrm{iso}$ stands for the isocurvature pressure perturbation defined as $\delta P_\mathrm{iso}\equiv c^2_\mathrm{s}\bar{\rho}_\mathrm{m}S$, where $S$ denotes the isocurvature perturbation, associated with the PBH density fluctuations introduced in \Sec{sec:P_Psi_PBH} sourcing as we will see later the induced GW signal studied in this work. The expressions for $\rho^{(1)}_\mathrm{f(R)}$ and $p^{(1)}_\mathrm{tot}$ can be found in~\cite{Zhou:2024doz}. From the above equations, we can see that the total energy and pressure densities together with their perturbations receive contributions from both the matter content of the Universe as well as from the curvature-induced effective fluid. Since we consider power-law type $f(R)$-driven inflationary scenarios~\cite{Odintsov:2025eiv}, where one can achieve a smooth transition to the post-inflationary radiation-dominated $\Lambda$CDM era~\cite{Fazzari:2025nfr}, we neglect the effect of $f(R)$ gravity at the level of $c^2_\mathrm{s,tot}$ and $w_\mathrm{tot}$ considering that, in a PBH-matter dominated era, $c^2_\mathrm{s,tot}\simeq w_\mathrm{tot} \simeq 0 $ as a first approximation.

As one may see from \Eq{eq:Phi_Psi} and \Eq{eq:Psi_minus_Phi}, the scalar metric perturbations $\Phi^{(1)}$ and $\Psi^{(1)}$ are in principle different. One then should express $\Phi^{(1)}$ as a function of $\Psi^{(1)}$ and derive from \Eq{eq:Phi_Psi} a differential equation for $\Psi^{(1)}$. Doing so, from \Eq{eq:Psi_minus_Phi}, we obtain
\beq\label{eq:Psi=Phi+R}
\Psi^{(1)}  = \Phi^{(1)} + \frac{F^{(
0)}_R}{F^{(
0)}}R^{(1)}
\eeq
where we have written $F^{(1)}$ as $F^{(1)} \equiv F^{(
0)}_RR^{(1)}$ and $R^{(1)}$ is the first order perturbed Ricci scalar reading as~\cite{Capozziello:2011et}
\beq\label{eq:Ricci_perturbed}
R^{(1)} = -\frac{2}{a^2}\Bigl[6(\mathcal{H}^\prime +\mathcal{H}^2)\Phi^{(1)} + 3\mathcal{H}\left(\Phi^{(1)\prime}  + 4\Psi^{(1)\prime}\right) + 3\Psi^{(1)\prime\prime} + \Delta \Phi^{(1)} - 2
\Delta\Psi^{(1)}\Bigr].
\eeq
At this point, one should note, developing in $k$ space the Fourier modes of $\Psi^{(1)}$ and $\Phi^{(1)}$, namely $\Psi^{(1)}_\boldmathsymbol{k}$ and $\Phi^{(1)}_\boldmathsymbol{k} $ respectively, that they can be written 
in terms of their transfer functions $T_\Psi$ and $T_\Phi$ as
\begin{align}
\Psi^{(1)}_\boldmathsymbol{k} & = T_\Psi(k\eta)\psi_\boldmathsymbol{k} \\
\Phi^{(1)}_\boldmathsymbol{k} & = T_\Phi(k\eta)\phi_\boldmathsymbol{k},
\end{align}
where $\psi_\boldmathsymbol{k}$ and $\phi_\boldmathsymbol{k}$ are the primordial values of $\Psi^{(1)}_\mathbf{k}$ and $\Phi^{(1)}_\mathbf{k}$ at an initial reference time, here the PBH domination time. Let us then derive below the evolution of $\Psi^{(1)}_\boldmathsymbol{k}$  and $\Phi^{(1)}_\boldmathsymbol{k}$ during MD era, considering separately super-horizon ($k\ll aH$) and sub-horizon ($k\gg aH$) scales.

\subsection{Super-horizon scales}\label{sec:scalar_prt_super_horizon}
In the super-horizon regime, i.e. when $k\ll aH$, one can neglect the $\Delta \Psi^{(1)}$ and $\Delta \Phi^{(1)}$ terms and write \Eq{eq:Phi_Psi} and \Eq{eq:Psi_minus_Phi} during a PBH-dominated era, where $w_\mathrm{tot}=c^2_\mathrm{tot}\simeq 0$, in terms of the transfer functions as $T_\Psi(k\eta)$ and $T_\Phi(k\eta)$ as
\begin{align}
&T^{\prime\prime}_\Psi + \mathcal{H}
\left[T_\Phi^{\prime}+2T_\Psi^{\prime}\right] = 0\label{eq:Phi_Psi_super} \\
&T_\Psi - T_\Phi = -\frac{2F^{(0)}_R}{a^2F^{(0)}}\left[3\mathcal{H}^2T_\Phi^{} + 6\mathcal{H}T_\Psi^{\prime}\right]\label{eq:Psi_minus_Phi_super},
\end{align}
where we plugged \Eq{eq:Ricci_perturbed} in \Eq{eq:Psi=Phi+R}.

\begin{itemize}
%%%%%%%%%%%%%%%%%%%%%%%%%%%%%%%%%%%%%%%%%%%%%%%
\item \textbf{$R^2$ gravity}

In the case of $R^{2}$ Starobinsky gravity, one has that the $f(R)$ function reads as
\beq
f(R) = R + \frac{R^2}{6M^2_\mathrm{s}}
\eeq
where $M_\mathrm{s} \simeq 10^{-5}\Mp$, fixed by the constraints of the Planck data on Starobinsky inflation~\cite{Planck:2018jri}. 
One  then can express from \Eq{eq:Psi_minus_Phi_super} $T_\Phi$ in terms of  $T_\Psi$ and get from \Eq{eq:Phi_Psi_super} a differential equation for $T_\Psi$ reading as
\beq\label{eq:Psi_ODE_super_R_2}
T_\Psi^{\prime\prime}\left[1 - 6p(\eta)\mathcal{H}^2\right]+ \mathcal{H}\left[1 - (6p(\eta)\mathcal{H})'\right]T_\Psi^{\prime} = 0,
\eeq
where $p(\eta)\equiv -\frac{2F^{(0)}_R}{a^2F^{(0)}}$. To extract \Eq{eq:Psi_ODE_super_R_2}, we accounted for the fact that for our $R^2$ gravity $p\mathcal{H}^2\ll 1$. Considering then a scale factor during the PBH dominated era where $a = a_\mathrm{d}\left(\frac{\eta}{\eta_\mathrm{d}}\right)^2$, with $a_\mathrm{d} = \frac{a_\mathrm{f}}{\Omega_\mathrm{PBH,f}}$, $\eta_\mathrm{d} = \frac{2}{a_\mathrm{d}H_\mathrm{d}}$, $H_\mathrm{d} = H_\mathrm{f}\Omega^2_\mathrm{PBH,f}$ and $a_\mathrm{f}$ given by \Eq{eq:scale_factor_PBH_formation}, one obtains that \Eq{eq:Psi_ODE_super_R_2} can be recast as
\beq\label{eq:Psi_ODE_super_R_3}
T_\Psi^{\prime\prime}\left(1+\frac{\lambda_1 M^3_\mathrm{PBH}}{\Mp^7M^2_s}\frac{1}{\eta^6}\right) + \frac{2}{\eta}\left(1 + \frac{\lambda_2 M^3_\mathrm{PBH}}{\Mp^7M^2_s}\frac{1}{\eta^6}\right)T_\Psi^{\prime} = 0,
\eeq
where $\lambda_2 = 5\lambda_1/2$ with $\lambda_1 = 2.5\times 10^{4}\frac{1}{g_\mathrm{eff}\Omega^{(0)3/2}_r}\left(\frac{\Mp^4}{\rho_0}\right)^{3/2}$. For $\g_\mathrm{eff} = 103$ and $\rho_0\simeq 10^{-120}\Mp^4$, one gets that $\lambda_1 = 8.5\times 10^{188}$. Imposing that $T_\Psi$ approaches $1$ in the past, i.e. $T_\Psi(k\eta\rightarrow 0) = 1$, one obtains that \Eq{eq:Psi_ODE_super_R_3} accepts an analytic solution reading as $T_\Psi(k\eta \ll 1) = 1$. Then, from \Eq{eq:Psi_minus_Phi_super}, we get that $T_\Psi/T_\Phi - 1  = - 3p \mathcal{H}^2\ll 1$. We checked also numerically that if we lower the value of $M_\mathrm{s}$ up to $10^{-15}\Mp$, $3p\mathcal{H}^2$ continued to remain far less than unity. We conclude thus that like in GR $T_\Psi(k\eta\ll 1) \simeq T_\Phi(k\eta\ll 1) = 1$, being conserved on super-horizon scales.

%%%%%%%%%%%%%%%%%%%%%%%%%%%%%%%%%%%%%%%%%%%%%%%
\item \textbf{$R^{1+\epsilon}$ gravity}

In the case of $R^{1+\epsilon}$ gravity, one can write the $f(R)$ function as
\beq
f(R) = M^2_\mathrm{s}\frac{R^{1+\epsilon}}{ M^{2(1+\epsilon)}_\mathrm{s}},
\eeq
where $M_\mathrm{s}$ is a free mass parameter of the theory and $\epsilon\ll 1$,  so that  $R^{1+\epsilon}$ gravity   deviates smoothly from GR.
This can be seen also by parameterizing the theory as 
\beq
R^{1+\epsilon} =R\cdot  R^{\epsilon}\simeq R\left[1+(\log R)\epsilon+\mathcal{O}(\epsilon^2)\right]\simeq R+\epsilon R \log R\,,
\eeq
where the correction with respect to GR is evident. It is worth noticing that this kind of logarithmic corrections assume a relevant role in problems of regularization and renormalization of the gravitational field (up to one-loop level)~\cite{Birrell:1982ix, Odintsov:2024lid, Acquaviva:2022bju, Dixit:2024grt, Kruglov:2023odm, Buchbinder:1992gdx}.

Following then the same reasoning as before,  we obtain a differential equation for $T_\Psi$ on super-horizon scales, i.e. $k\eta \ll 1$, and in the limit $\epsilon \ll 1$, reading as
\beq\label{eq:Psi_ODE_super}
T_\Psi^{\prime\prime}\left(1+11\epsilon\right) + \frac{7}{\eta}T_\Psi^{\prime} = 0
\eeq
One can then write the solution of \Eq{eq:Psi_ODE_super} as 
\beq
T_\Psi(x\ll 1, \epsilon\ll 1) \simeq c_1 -\frac{c_2}{6x^6} - \frac{77c_2(-5+6\ln x)\epsilon}{36x^6},
\eeq
where $x = k\eta$. Imposing that $T_\Psi$ approaches $1$ in the past, i.e. $T_\Psi(x\rightarrow 0) = 1$, one gets that $c_1 = 1$ and $c_2=0$. At the end, we straightforwardly get that $T_\Psi(x\ll 1) = 1$. Regarding $T_\Phi$, one can show from \Eq{eq:Psi_minus_Phi_super} that $T_\Psi - T_\Phi \simeq 2\epsilon T_\Phi\ll 1$, since $\epsilon\ll 1$. Consequently, on super-horizon scales one obtains that 
\beq\label{eq:Phi_Psi_solution_super}
T_\Psi(k\eta \ll 1 )\simeq T_\Phi(k\eta \ll 1 ) \simeq 1.
\eeq
We conclude then again that  $T_\Psi$ and $T_\Phi$ are conserved on super-horizon scales like in GR.

%%%%%%%%%%%%%
\end{itemize}

\subsection{Sub-horizon scales}\label{sec:scalar_prt_sub_horizon}
Working in the sub-horizon regime 
\Eq{eq:Ricci_perturbed} will be recast as~\cite{Tsujikawa:2007gd}
\beq\label{eq:Ricci_perturbed_subhorizon}
R^{(1)} \simeq -\frac{2}{a^2}\left(\Delta \Phi^{(1)}-2\Delta \Psi^{(1)}\right).
\eeq
Plugging thus \Eq{eq:Ricci_perturbed_subhorizon} into \Eq{eq:Psi=Phi+R} and going to the Fourier space one gets that 
\beq\label{eq:Phi_vs_Phi}
\Phi^{(1)}_\boldmathsymbol{k} = \frac{1+4\frac{k^2}{a^2}\frac{F^{(
0)}_R}{F^{(
0)}}}{1+2\frac{k^2}{a^2}\frac{F^{(
0)}_R}{F^{(
0)}}}\Psi^{(1)}_\boldmathsymbol{k}.
\eeq

For the case of a matter era driven by PBHs, one has $c^2_\mathrm{s,tot} \simeq w_\mathrm{tot} = 0 $ since PBHs behave effectively as presureless dust. At the end, inserting \Eq{eq:Phi_vs_Phi} into \Eq{eq:Phi_Psi} and working in the Fourier space, one gets for $T_\Psi$ that
\beq\label{eq:Psi_eMD}
T_\Psi^{\prime\prime} + \mathcal{H}\left[2+\frac{1+4g(\eta,k)}{1+2g(\eta,k)}\right]T_\Psi^{\prime} + \left[\mathcal{H}\left(\frac{1+4g(\eta,k)}{1+2g(\eta,k)}\right)^\prime - k^2\frac{g(\eta,k)}{1+2g(\eta,k)}\right]T_\Psi = 0,
\eeq
where $g(\eta,k)\equiv \frac{k^2}{a^2}\frac{F^{(
0)}_R}{F^{(
0)}} $. 
Before going into investigating our particular $f(R)$ gravity models, it is useful to define based on \Eq{eq:Phi_vs_Phi} the ratio $\lambda \equiv \Phi^{(1)}_\boldmathsymbol{k}/\Psi^{(1)}_\boldmathsymbol{k}$  as a measure of the difference between $\Phi^{(1)}_\boldmathsymbol{k}$ and $\Psi^{(1)}_\boldmathsymbol{k}$, usually quoted as gravitational slip. Namely, one has from \Eq{eq:Phi_vs_Phi} that 
\beq
\lambda(k,\eta) = \frac{1+4\frac{k^2}{a^2}\frac{F^{(
0)}_R}{F^{(
0)}}}{1+2\frac{k^2}{a^2}\frac{F^{(
0)}_R}{F^{(
0)}}}.
\eeq

\begin{figure}[ht]
\begin{center}
\includegraphics[width = 0.85\textwidth]{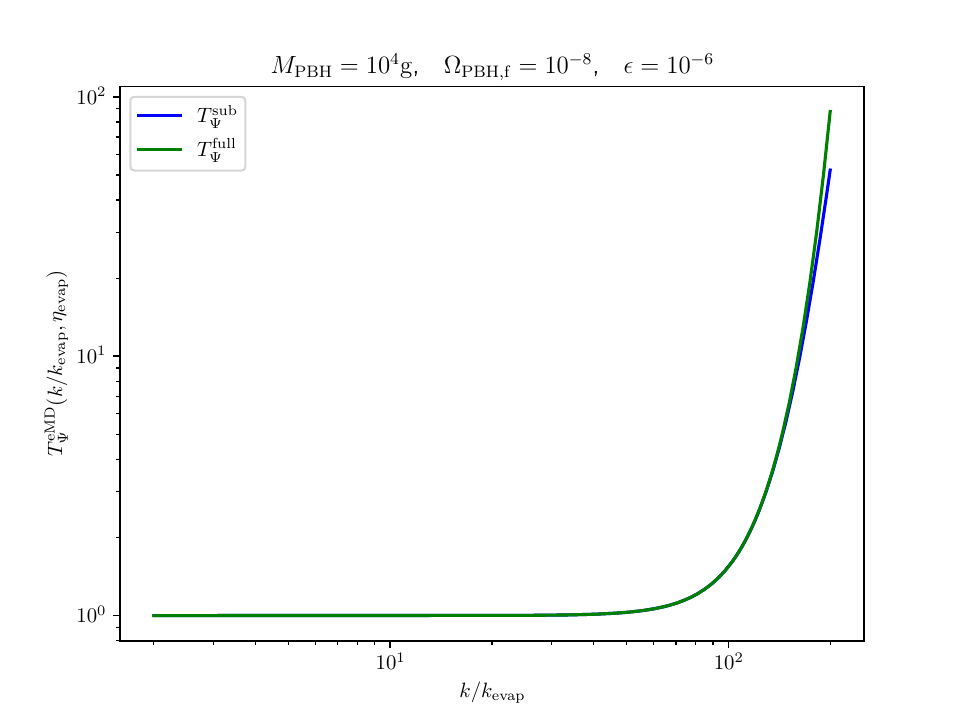}
\caption{{The transfer function $T_\Psi(x,\epsilon)$ at the end of the PBH-driven eMD era for $M_\mathrm{PBH} = 10^4\mathrm{g}$, $\Omega_\mathrm{PBH,f} = 10^{-8}$ and $\epsilon = 10^{-6}$ as a function of $k/k_\mathrm{evap}$ (green curve) superimposed with $T_\Psi(x,\epsilon)$ at sub-horizon (blue curve) and super-horizon scales (red curve) scales.}}
\label{fig:T_Psi_eMD}
\end{center}
\end{figure}

\begin{figure}[ht]
\begin{center}
\includegraphics[width = 0.85\textwidth]{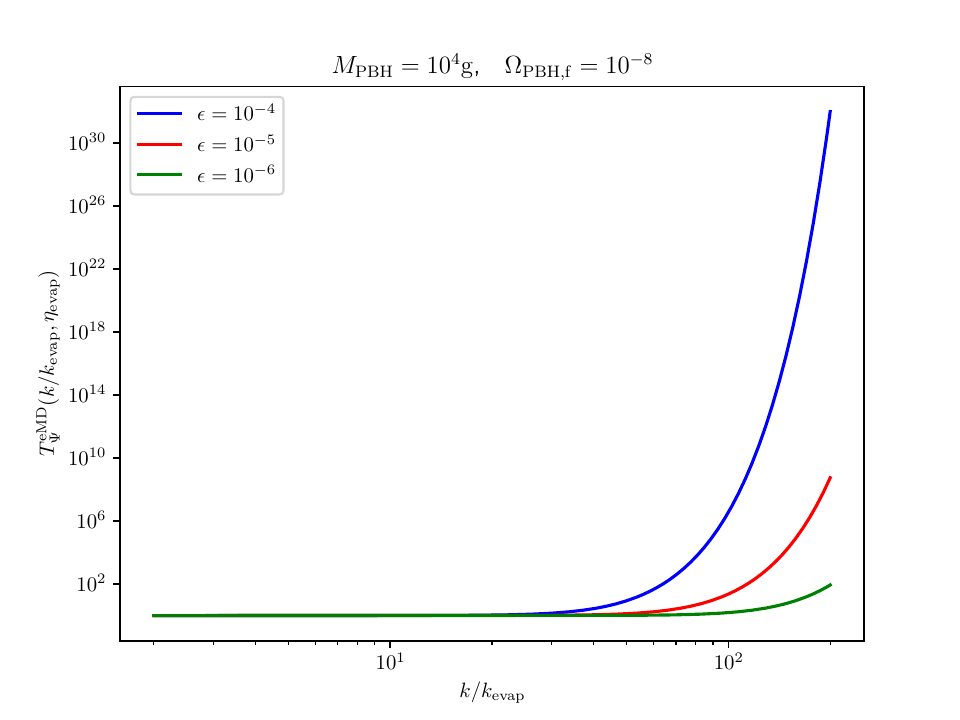}
\caption{{The transfer function $T_\Psi(x,\epsilon)$ at the end of the PBH-driven eMD era for $M_\mathrm{PBH} = 10^5\mathrm{g}$ and $\Omega_\mathrm{PBH,f} = 10^{-7}$ and for different values of the parameter $\epsilon$.}}
\label{fig:T_Psi_eMD_vs_epsilon}
\end{center}
\end{figure}

\begin{itemize}
    \item \textbf{$R^2$ gravity}
    
In the case of $R^2$ gravity, we checked numerically that the ratio $\lambda(k,\eta)=\Phi^{(1)}_\boldmathsymbol{k}/\Psi^{(1)}_\boldmathsymbol{k}$ for different values of the PBH mass $M_\mathrm{PBH}$, the initial PBH abundance $\Omega_\mathrm{PBH,f}$ and the mass parameter $M_\mathrm{s}$, approaches unity, namely $\lambda(k,\eta)\rightarrow 1$, like in the case of GR. One then naively concludes that $T_\Psi (k\eta \gg 1) \simeq 1$ in $R^2$ gravity.

%%%%%%%%%%%%%%%%%%%%%%%%%%%%%%%%%%%%%%%%%%%%%%%%%%%%%%%%%%%%%%%%%%%%%%%%%%%%%%%%%%%%%%%%%%
\item \textbf{$R^{1+\epsilon}$ gravity}

However, for $R^{1+\epsilon}$ gravity, the ratio $\Phi^{(1)}_\boldmathsymbol{k}/\Psi^{(1)}_\boldmathsymbol{k}$ is not necessarily unity. In particular, for small values of $k$ close to $k_\mathrm{evap}$, $\Phi^{(1)}_\boldmathsymbol{k}/\Psi^{(1)}_\boldmathsymbol{k}$ approaches unity whereas for $k\to k_\mathrm{UV}$, we obtain $\Phi^{(1)}/\Psi^{(1)}\rightarrow 2$. The results do not change drastically by varying the PBH mass $M_\mathrm{PBH}$, the PBH abundance $\Omega_\mathrm{PBH,f}$ and the mass parameter $M_\mathrm{s}$. One then should treat for this particular $f(R)$ gravity model $\Phi^{(1)}$ and $\Psi^{(1)}$ as different.

Doing so, one should solve \Eq{eq:Psi_eMD} for the dynamical evolution of $T_\Psi$. Making a Taylor expansion of $\frac{1+4g(\eta,k)}{1+2g(\eta,k)}$ and $\frac{2g(\eta,k)}{1+2g(\eta,k)}$ for small $\epsilon$, i.e. $\epsilon \ll 1$, one can write \Eq{eq:Psi_eMD} as follows:
\beq
\frac{\mathrm{d}^2T_\Psi}{\mathrm{d}x^2}+ \frac{2}{x}\left(3+\frac{x^2\epsilon}{6}\right)\frac{\mathrm{d}T_\Psi}{\mathrm{d}x} + \frac{\epsilon}{6}\left(4-\frac{x^2}{2}\right)T_\Psi = 0,
\eeq
where $x = k\eta$. The above equation has an analytic solution which reads as
\beq\label{eq:Psi_eMD_R_1+epsilon_full}
T_\Psi(x,\epsilon) = e^{-\frac{x^2}{12}\left(\epsilon + \sqrt{\epsilon(3+\epsilon)}\right)} \; _1F_1\left(\frac{7}{4}+\frac{3\epsilon}{4\sqrt{\epsilon(3+\epsilon)}},\frac{7}{2},\frac{x^2}{6}\sqrt{\epsilon(3+\epsilon)}\right),
\eeq
with $\lim_{\epsilon\rightarrow 0}T_\Psi(x,\epsilon) = 1$, recovering thus the GR regime in the limit $\epsilon\rightarrow 0$. Deeply in the sub-horizon regime, i.e. $k\eta\gg 1$, one gets that
\beq\label{eq:Psi_eMD_R_1+epsilon}
T_\Psi(k\eta\gg 1,\epsilon) \simeq 1 + \frac{x^4\epsilon}{216}+\frac{x^8\epsilon^2}{134784} + 
 \frac{x^{12}\epsilon^3}{164975616}.
\eeq
In the limit $\epsilon\rightarrow 0$, we recover the GR result, i.e. $\lim_{\epsilon\rightarrow 0}T_\Psi(x\gg 1,\epsilon) = 1$.

\end{itemize}
In \Fig{fig:T_Psi_eMD}, we show the transfer function $T_\Psi(x,\epsilon)$ at the end of the PBH-eMD era (green curve) superimposed with $T_\Psi(x,\epsilon)$ at sub-horizon (blue curve) while in \Fig{fig:T_Psi_eMD_vs_epsilon} we see how $T_\Psi(x,\epsilon)$ behavies by varying the parameter $\epsilon$.  As one can infer from \Fig{fig:T_Psi_eMD}, $T_\Psi(x,\epsilon)$ increases very abruptly as one goes to small scales experiencing more time within the PBH-eMD era. In contrast with GR, where $\Psi$ does not grow during a MD era, i.e. $T^\mathrm{GR}_\Psi = 1$, in $R^{1+\epsilon}$ gravity, scalar perturbations $\Psi$ grow exponentially as they stay more within the PBH-eMD era. In \Fig{fig:T_Psi_eMD_vs_epsilon} we show as well how $T_\Psi$ behaves in response of the parameter $\epsilon$. Interestingly enough, as we increase the value of $\epsilon$, i.e. departing from the GR limit, more and more larger scales from the small $k$ region become non-linear entering earlier the non-perturbative regime.

This rapid growth of the scalar perturbations was already discussed in ~\cite{Carloni:2007yv,Ananda:2008tx} where it was shown that for $R^n$ gravity with $n\sim 1$, the behaviour of the matter perturbations can be drastically different as compared to GR leading potentially to an instability of the theory. This can be explained from the fact that even small deviations from linearity in the action, i.e. $R^{1^{\pm}}$, result in a change of order (from two to four) of the matter perturbations equations. Since the study of how to cure such instabilities is beyond the scope of our work, we impose a UV cut-off scale $k_\mathrm{c}<k_\mathrm{UV}$ below which we enter the non-linear regime, i.e. $\mathcal{P}_\Psi(k>k_\mathrm{c},\eta_\mathrm{evap})>1$, where one cannot perform an analysis with standard perturbation theory techniques [See also the discussion after \Eq{eq:k_f}.]. Note that similar exponential growth of perturbations was found as well within the context of GR during a preheating era driven by a $\lambda\phi^4$ inflationary potential~\cite{Jedamzik:2010dq}, where one expects rapid structure formation. We need to note as well that this interesting feature can have a significant impact for early structure formation during eMD in $R^{1+\epsilon}$ and is worth being investigated in a separate work, since it will require the performance of high-cost numerical simulations~\cite{Jedamzik:2010dq,Barenboim:2013gya,Eggemeier:2020zeg}.

\subsection{The transition to the radiation-dominated era}
Going from the PBH-driven eMD era to the late RD (lRD) era, one should require the continuity of the Bardeen gravitational potentials and their first time derivatives at the evaporation time, i.e. at the transition time from the eMD to the lRD era~\footnote{One should note here that the matching of the Bardeen gravitational potentials and their first time derivatives at evaporation time should be strictly performed in the synchronous comoving gauge, where all PBHs evaporate at the same time. In the Appendix \ref{app:Synchronous_gauge}, we show that the continuity of $\Psi$ and $\Psi'$ in the synchronous comoving gauge implies their continuity in the Newtonian gauge as well.}.

\begin{itemize}
    \item \textbf{$R^2$ gravity}
    
For the case of Starobinsky $R^2$ gravity, one has that during the eMD era, $T_\Psi= T_\Phi = 1$, as we showed in the previous sections, whereas during the RD era, during which the Ricci scalar $R$ is zero, i.e. there is no anisotropic stress,  $\Phi^{(1)} = \Psi^{(1)}$ and one recovers the GR solutions for the transfer function of the Bardeen gravitational potentials, being recast as~\cite{Domenech:2020ssp}
\beq\label{eq:GR_T_Phi_Psi}
T_\Phi^{R^2,\mathrm{lRD}}= T_\Psi^{R^2,\mathrm{lRD}} = \frac{1}{c_\mathrm{s}k\eta}\left[C_{1,R^2}j_1(c_\mathrm{s}k\eta) + C_{2,R^2}y_1(c_\mathrm{s}k\eta)\right],
\eeq
where $j_i$ and $y_i$ are the spherical Bessel functions of order $i$. 
Requiring thus that $T_\Psi^{R^2,\mathrm{lRD}} (k\eta_\mathrm{evap})= T_\Psi^{R^2,\mathrm{eMD}}(k\eta_\mathrm{evap})$ and  $T_\Psi^{R^2,\mathrm{lRD},\prime}(k\eta_\mathrm{evap}) = T_\Psi^{R^2,\mathrm{eMD},\prime}(k\eta_\mathrm{evap})$, one gets that 
\begin{align}
    C_{1,R^2} & = c^2_\mathrm{s}k^2\eta^2_\mathrm{evap}\left[\cos(c_\mathrm{s}k\eta_\mathrm{evap}) + 3j_1(c_\mathrm{s}k\eta_\mathrm{evap})\right]  \\
    C_{2,R^2} & = -c^2_\mathrm{s}k^2\eta^2_\mathrm{evap}\left[\sin(c_\mathrm{s}k\eta_\mathrm{evap}) - 3y_1(c_\mathrm{s}k\eta_\mathrm{evap})\right].  
\end{align}
Note that $T_\Psi^{R^2,\mathrm{eMD}}=T_\Phi^{R^2,\mathrm{eMD}}\simeq 1$ as shown in \Sec{sec:scalar_prt_super_horizon} and in \Sec{sec:scalar_prt_sub_horizon}.

\item \textbf{$R^{1+\epsilon}$ gravity} 
 
For the case of $R^{1+\epsilon}$ gravity, $\Psi^{(1)}$ and $\Phi^{(1)}$ are the same on superhorizon scales but are different on sub-horizon scales [See \Eq{eq:Phi_vs_Phi}] during the PBH-dominated era. However, during the late RD era, the Ricci scalar vanishes and so does the anisotropic stress leading to $\Phi^{(1)} = \Psi^{(1)}$ and to the GR recovery. Focusing thus on sub-horizon scales, which are the scales of interest for us, and requiring the continuity of the Bardeen gravitational potentials and their first time derivatives at the evaporation time, the transfer functions \eqref{eq:GR_T_Phi_Psi} during the late RD era will read as 
\begin{align}\label{eq:T_Psi_RD_FR}
T_\Phi^{R^{1+\epsilon},\mathrm{lRD}} =T_\Psi^{R^{1+\epsilon},\mathrm{lRD}} = \frac{-1}{(c_\mathrm{s}k\eta)^2}\left[C^\Psi_{1,R^{1+\epsilon}}\cos(c_\mathrm{s}k\eta)+ C^\Psi_{2,R^{1+\epsilon}}\sin(c_\mathrm{s}k\eta)\right],
\end{align}
with 
\begin{align}
    \begin{split}
    C^\Psi_{1,R^{1+\epsilon}} & = \frac{T_\Psi^{R^{1+\epsilon},\mathrm{eMD}}(k\eta_\mathrm{evap})}{6}c_\mathrm{s}k\eta_\mathrm{evap}\Biggl\{-c_\mathrm{s}k\eta_\mathrm{evap}\cos(c_\mathrm{s}k\eta_\mathrm{evap}) \\ & + 
    2\sin(c_\mathrm{s}k\eta_\mathrm{evap}) \Bigl[2291328 k^4 \eta^4_\mathrm{evap}\epsilon  + 
 6120 k^8 \eta^8_\mathrm{evap}\epsilon^2 \\ & + 
 7 k^{12} \eta^{12}_\mathrm{evap}\epsilon^3 + 
 164975616
 \Bigr]\\ & /\Bigl[763776 k^4 \eta^4_\mathrm{evap}\epsilon + 
 1224 k^8 \eta^8_\mathrm{evap}\epsilon^2 + 
 k^{12} \eta^{12}_\mathrm{evap}\epsilon^3 + 
164975616
\Bigr]\Biggr\}
    \end{split}
    \\
    \begin{split}
    C^\Psi_{2,R^{1+\epsilon}} & = \frac{T_\Psi^{R^{1+\epsilon},\mathrm{eMD}}(k\eta_\mathrm{evap})}{6}c_\mathrm{s}k\eta_\mathrm{evap}\Biggl\{-c_\mathrm{s}k\eta_\mathrm{evap}\sin(c_\mathrm{s}k\eta_\mathrm{evap}) \\ & - 
    2\cos(c_\mathrm{s}k\eta_\mathrm{evap}) \Bigl[2291328 k^4 \eta^4_\mathrm{evap}\epsilon  + 
 6120 k^8 \eta^8_\mathrm{evap}\epsilon^2 \\ & + 
 7 k^{12} \eta^{12}_\mathrm{evap}\epsilon^3 + 
164975616
\Bigr]\\ & /\Bigl[763776 k^4 \eta^4_\mathrm{evap}\epsilon + 
 1224 k^8 \eta^8_\mathrm{evap}\epsilon^2 \ + 
 k^{12} \eta^{12}_\mathrm{evap}\epsilon^3 + 
 164975616
\Bigr]\Biggr\}
    \end{split}, 
\end{align}
where $T_\Psi^{R^{1+\epsilon},\mathrm{eMD}}$ is given by \Eq{eq:Psi_eMD_R_1+epsilon}.
\end{itemize}
\subsection{The primordial black hole gravitational potential in \texorpdfstring{$f(R)$}{f(R)} gravity.}\label{sec:Psi_Phi_PBH_f_R_gravity}
Having seen above how the gravitational potentials $\Psi^{(1)}$ and $\Phi^{(1)}$ evolve during the transition from the PBH-dominated to the late RD era let us derive here the power spectrum for $\Psi^{(1)}$ and $\Phi^{(1)}$, i.e. $\mathcal{P}_\Psi$ and $\mathcal{P}_\Phi$ respectively, being the source of the SIGW signal we will study in the following.

Doing so, like in the case of GR, one should connect $\mathcal{P}_\Psi$ and $\mathcal{P}_\Phi$  with the PBH matter power spectrum $\mathcal{P}_{\delta_\mathrm{PBH}}$ [See \Sec{sec:P_Psi_PBH}], since the PBH matter fluctuations are the ones who are sourcing the scalar metric perturbations $\Psi^{(1)}$ and $\Phi^{(1)}$ in our physical setup.  
In order to proceed, we introduce the uniform-energy density curvature perturbation of a fluid, $\zeta$,being associated to the gravitational potential $\Phi^{(1)}$ and the respective energy density perturbation as \cite{Wands:2000dp}:
\begin{equation}
    \zeta \equiv -\Phi^{(1)} - \mathcal{H} \frac{\delta \rho}{\bar{\rho}'}.
\end{equation}
Accounting thus for the continuity equation $\bar{\rho} ' = - 3\mathcal{H} (\bar{\rho} + \bar{p})$, one gets that $\zeta $ can be recast as
\begin{equation}
  \zeta \equiv -\Phi^{(1)} + \frac{\delta}{3(1+w)}, \label{z}
\end{equation}
where $w$ is the fluid equation-of-state parameter. In our case, we have a three fluid system, consisting in form of PBH matter, radiation and the $f(R)$ curvature-induced fluid with energy density and pressure given by \Eq{reff} and \Eq{peff}. Since the energy-momentum tensors of radiation, PBH-matter and the $f(R)$ curvature-induced fluid are separately conserved~\footnote{We need to
stress here that in $f(R)$ gravity models, $\zeta_\mathrm{r}$ and $\zeta_\mathrm{PBH}$ are not generically conserved. However, in metric gravity theories where the total matter-stress energy tensor
is conserved, $\zeta_\mathrm{r}$ and $\zeta_\mathrm{PBH}$ are conserved as well~\cite{Wands:2000dp}. This is the case in our
study, since working within the ``effective curvature-induced fluid approach"~\cite{Sotiriou:2008rp}, $f(R)$ gravity can be seen as GR with a matter content consisting of the ordinary matter one (PBHs and radiation in our case) plus the ``effective curvature-induced" fluid characterised by its own energy density and pressure. [See the Appendices A and C of~\cite{Papanikolaou:2021uhe} for more details].},  we can use (\ref{z}) for
$\zeta_\mathrm{r}$, $\zeta_\mathrm{PBH}$ and $\zeta_{f(R)}$ and get that :
\beq\label{eq:zeta_r}
\zeta_\mathrm{r}=-\Phi^{(1)}+\frac{1}{4} \delta_\mathrm{r},
\eeq
\beq\label{eq:zeta_PBH}
\zeta_\mathrm{PBH}=-\Phi^{(1)}+\frac{1}{3} \delta_\mathrm{PBH},
\eeq
\beq\label{eq:zeta_f_R}
\zeta_{f(R)}=-\Phi^{(1)}+\frac{1}{3(1+w_{f(R)}} \delta_{f(R)},
\eeq
while the total curvature perturbation $\zeta$ will be written as
\beq
\zeta=-\Phi^{(1)}+\frac {\delta_{\mathrm{tot}}}{3(1+w_\mathrm{tot})} = \frac{\frac{4}{3}\bar{\rho}_\mathrm{r}\zeta_\mathrm{r} + \bar{\rho}_\mathrm{PBH}\zeta_\mathrm{PBH} + (1 + w_\mathrm{f(R)})\bar{\rho}_\mathrm{f(R)} \zeta_\mathrm{f(R)}}{\frac{4}{3}\bar{\rho}_\mathrm{r} + \bar{\rho}_\mathrm{PBH} + (1 + w_\mathrm{f(R)})\bar{\rho}_\mathrm{f(R)}} \label{zeta}.
\eeq

Here, it is important to highlight that we consider PBHs produced during the RD era after the end of inflation. It is thus expected that at the background level, the energy contribution from the curvature-induced fluid will be negligible compared to the contribution of radiation and PBH matter, i.e. $\bar{\rho}_\mathrm{r}/\bar{\rho}_\mathrm{f(R)} \gg 1$ and $\bar{\rho}_\mathrm{PBH}/\bar{\rho}_\mathrm{f(R)} \gg 1$. Furthermore, on the small scales we focus on, associated with the PBH formation scale, the dominant contribution to $\zeta$ during the PBH dominated era and the subsequent late RD era after PBH evaporation, will be due to $\zeta_\mathrm{PBH}$ and $\zeta_\mathrm{r}$. One then can neglect $(1 + w_\mathrm{f(R)})\bar{\rho}_\mathrm{f(R)} \zeta_\mathrm{f(R)}$ and $(1 + w_\mathrm{f(R)})\bar{\rho}_\mathrm{f(R)}$ from the numerator and the denominator of \Eq{zeta} respectively. Therefore, $\zeta$ will read as 
 \begin{equation}\label{eq:zeta_r_PBH}
     \zeta = \frac{4}{4+3s}\zeta_\mathrm{r}+\frac{3s}{4+3s}\zeta_\mathrm{PBH} ,     
 \end{equation}
with $ s\equiv \frac{a}{a_\mathrm{d}}$, 
and $a_\mathrm{d}$ standing for the scale factor $a$ at the time of PBH domination. 

Having thus effectively a two-fluid system, as one can see from \Eq{eq:zeta_r_PBH}, we can introduce the isocurvature perturbation defined as:
\beq\label{eq:S definition}
S = 3\left(\zeta_\mathrm{PBH}-\zeta_\mathrm{r}\right) =\delta_\mathrm{PBH} - 
\frac{3}{4} \delta_\mathrm{r}.
\eeq
During thus the PBH-driven eMD era, $\zeta\simeq \zeta_\mathrm{PBH} = 
\zeta_{\mathrm{r}}+S/3 \simeq S/3$. 

Focusing then on superhorizon scales, $\zeta_\mathrm{r}$ and  $\zeta_\mathrm{PBH}$ are 
conserved separately~\cite{Wands:2000dp}, so does the isocurvature perturbation 
$S$ and it can be calculated 
at the time of PBH domination $t_\mathrm{d}$.  Therefore, with $\zeta_\mathrm{r}$ at PBH domination time being negligible for the scales considered here, one can write from 
\Eq{eq:S definition} the isocurvature perturbation as $S=\delta_\mathrm{PBH}(t_\mathrm{d})$ with $\zeta$ being related to the PBH density perturbation as
\bea
\label{eq:zeta:delta:superH_GR}
\zeta\simeq \frac{1}{3} \delta_\mathrm{PBH}(t_\mathrm{d})\quad \mathrm{if}\quad 
k\ll \mathcal{H}\,. 
\eea

Finally, in order to relate the gravitational potentials $\Psi^{(1)}$ and $\Phi^{(1)}$ with $\delta_\mathrm{PBH}$, we exploit the fact that $\zeta\simeq -\mathcal{R}$ on superhorizon scales~\cite{Wands:2000dp}, where $\mathcal{R}$ is the comoving curvature 
perturbation defined by
\bea
\label{eq:zeta:Bardeen}
\mathcal{R}  = 
\frac{2}{3}\frac{{\Psi^{(1)}}^\prime/\mathcal{H}+\Phi^{(1)}}{1+w }+\Psi^{(1)}\, ,
\eea
and being related with $\Psi^{(1)}$ and $\Phi^{(1)}$. Since, on superhorizon scales $\Psi^{(1)} \simeq \Phi^{(1)} \simeq  \mathrm{constant\;in\;time}$ [See \Sec{sec:scalar_prt_super_horizon}],
one obtains straightforwardly that in the PBH-driven eMD era, where $w=0$, 
\bea
\label{eq:Phi:delta:superH00}
\Phi^{(1)} \simeq \Psi^{(1)} \simeq -\frac{1}{5} \delta_\mathrm{PBH}(t_\mathrm{d})\quad \mathrm{for}\quad 
k\ll \mathcal{H}.
\eea

Considering then sub-horizon scales, one can relate $\Psi^{(1)}$ and $\Phi^{(1)}$ with $\delta_\mathrm{PBH}$ by combining the Poisson equation and the M\'eszaros growth equation in $f(R)$ gravity [See also Appendix \ref{app:delta_PBH}] to write $\Psi^{(1)}$ and $\Phi^{(1)}$ as~\footnote{For more details on the basic steps to arrive at \Eq{eq:Psi_sub_PBH_gas_t_d} and \Eq{eq:Phi_sub_PBH_gas_t_d} the interested reader should go to Sec. III of~~\cite{Tsujikawa:2007gd} and the Sec. 3.3 of~\cite{Papanikolaou:2021uhe}.}
\begin{align}\label{eq:Psi_sub_PBH_gas_t_d}
  \Psi^{(1)} & \simeq -\frac{3}{2}\frac{\mathcal{H}^2}{k^2}\frac{1}{F}\frac{1+2\frac{k^2}{a^2}\frac{F^{(0)}_R}{F^{(0)}}}{1+3\frac{k^2}{a^2}\frac{F^{(0)}_R}{F^{(0)}}}\delta_\mathrm{PBH} \quad \mathrm{for}\quad 
k\gg \mathcal{H}, \\
   \Phi^{(1)} & \simeq -\frac{3}{2}\frac{\mathcal{H}^2}{k^2}\frac{1}{F}\frac{1+4\frac{k^2}{a^2}\frac{F^{(0)}_R}{F^{(0)}}}{1+3\frac{k^2}{a^2}\frac{F^{(0)}_R}{F^{(0)}}}\delta_\mathrm{PBH}\quad \mathrm{for}\quad 
k\gg \mathcal{H}\label{eq:Phi_sub_PBH_gas_t_d}.
\end{align}
Interpolating then between \Eq{eq:Phi:delta:superH00} and \Eq{eq:Psi_sub_PBH_gas_t_d}, \Eq{eq:Phi_sub_PBH_gas_t_d} and accounting for the PBH matter power spectrum \eqref{eq:ppoi} one can write $\mathcal{P}_\Psi$ and $\mathcal{P}_\Phi$ at the time of PBH domination as 
\begin{align}
    \mathcal{P}_\Psi(k,\eta_\mathrm{d}) & = \frac{k^3}{2\pi^2}P_\Psi(k) = \frac{2}{3\pi}\left(\frac{k}{k_\mathrm{UV}}\right)^3\left[5+\frac{2}{3}\left(\frac{k}
{\mathcal{H}_\mathrm{d}}\right)^2F^{(0)}_\mathrm{d} \left(\frac{1+ 3 \frac{k^2}{a^2_\mathrm{d}}\frac{F^{(0)}_{R_\mathrm{d}}}{F^{(0)}_\mathrm{d}}}{1+ 2 \frac{k^2}{a^2_\mathrm{d}} \frac{F^{(0)}_{R_\mathrm{d}}}{F^{(0)}_\mathrm{d}}}\right)\right]^{-2}\label{eq:P_Psi_eta_d} \\
 \mathcal{P}_\Phi(k,\eta_\mathrm{d}) & = \frac{k^3}{2\pi^2}P_\Phi(k) = \frac{2}{3\pi}\left(\frac{k}{k_\mathrm{UV}}\right)^3\left[5+\frac{2}{3}\left(\frac{k}
{\mathcal{H}_\mathrm{d}}\right)^2F^{(0)}_\mathrm{d} \left(\frac{1+ 3 \frac{k^2}{a^2_\mathrm{d}}\frac{F^{(0)}_{R_\mathrm{d}}}{F^{(0)}_\mathrm{d}}}{1+ 4 \frac{k^2}{a^2_\mathrm{d}} \frac{F^{(0)}_{R_\mathrm{d}}}{F^{(0)}_\mathrm{d}}}\right)\right]^{-2}\label{eq:P_Phi_eta_d}.
\end{align}

At this point, it is important to notice that that scales which have a time variation larger than the PBH evaporation rate $\Gamma\equiv - {\rm d}\ln M_{\rm PBH}/{\rm d}t$, i.e. $k/a\gg \Gamma$, are effectively suppressed by the non-zero pressure of the radiation fluid. This leads to the appearance of an extra suppression factor $S_G(k)$ in the expressions for $\mathcal{P}_\Psi(k,\eta)$ and $\mathcal{P}_\Phi(k,\eta)$ reading as~\cite{Inomata:2020lmk} 
\beq\label{eq:supfactor}
S_G(k)\equiv\left(\sqrt{\frac{2}{3}}\frac{k}{k_\mathrm{evap}}\right)^{-1/3}\,\,,
\eeq
where $k_{\rm evap}$ is the wavenumber of the mode crossing the Hubble horizon at the time of evaporation $\eta_{\rm evap}$.
At the end, the power spectra of $\Psi^{(1)}$ and $\Phi^{(1)}$ at the end of the PBH-driven eMD era will be recast as
\begin{align}\label{eq:Psi_Phi_sub_PBH_gas}
\mathcal{P}_\Psi(k,\eta_\mathrm{evap}) &=\tilde{\mathcal{P}}_\Psi(k,\eta_\mathrm{d})T^2_\Psi(k\eta_\mathrm{evap})\\
\mathcal{P}_\Phi(k,\eta_\mathrm{evap}) &= \tilde{\mathcal{P}}_\Phi(k,\eta_\mathrm{d})T^2_\Phi(k\eta_\mathrm{evap}),
\end{align}
where 
\begin{eqnarray}
\tilde{\mathcal{P}}_\Psi(k,\eta_\mathrm{d})& \equiv S^2_G(k) \mathcal{P}_\Psi(k,\eta_\mathrm{d}), \label{eq:P_Psi_eff} \\ 
\tilde{\mathcal{P}}_\Phi(k,\eta_\mathrm{d})&\equiv S^2_G(k) \mathcal{P}_\Phi(k,\eta_\mathrm{d}).\label{eq:P_Phi_eff}
\end{eqnarray}

\begin{figure}[ht]
\begin{center}
\includegraphics[width = 0.49\textwidth]{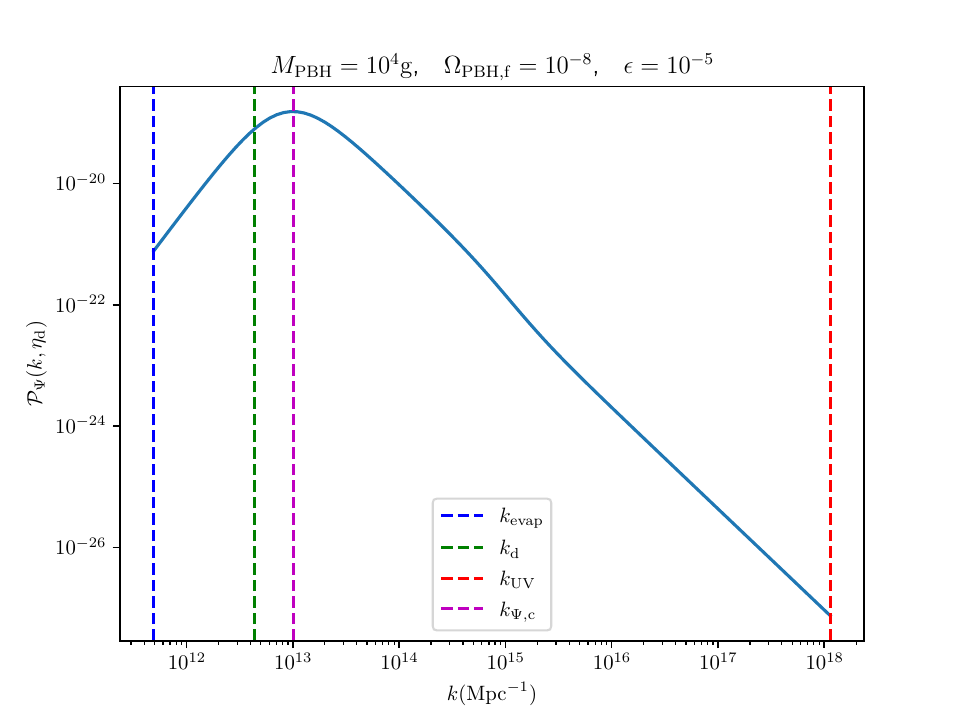}
\includegraphics[width = 0.49\textwidth]{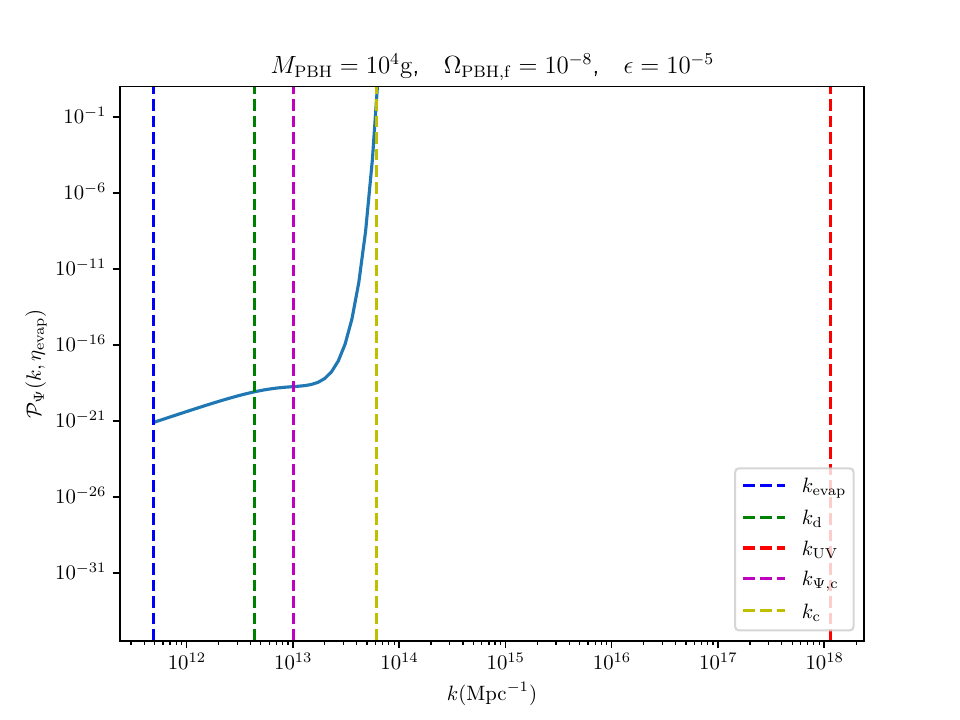}
\caption{{Left Panel: The scalar power spectrum $\mathcal{P}_\Psi$ at the onset of the PBH-eMD era. Right Panel: The scalar power spectrum $\mathcal{P}_\Psi$ at the end of the PBH-eMD era. In both panels we have fixed $M_\mathrm{PBH} = 10^4\mathrm{g}$, $\Omega_\mathrm{PBH,f}=10^{-8}$ and $\epsilon = 10^{-5}$. With the blue, green, red, magenta and yellow vertical dashed lines we show the characteristic scales $k_\mathrm{evap}$, $k_\mathrm{d}$, $k_\mathrm{UV}$, $k_{\Psi, \mathrm{c}}$ and $k_\mathrm{c}$ respectively.}}
\label{fig:P_Psi_d_vs_evap}
\end{center}
\end{figure}

In \Fig{fig:P_Psi_d_vs_evap}, we show the scalar power spectrum $\mathcal{P}_\Psi$ at the onset of the PBH-eMD era (left panel) and at the end of the PBH-eMD era (right panel). As it can be inferred from the left panel, one can identify four characteristic scales, namely the scales crossing the cosmological horizon at the onset and at the end of the PBH-eMD era, $k_\mathrm{d}$ and $k_\mathrm{evap}$ respectively, the UV cut-off scale $k_\mathrm{UV}$ below which PBH matter power spectrum $\mathcal{P}_{\delta_\mathrm{PBH}}(k)$ becomes non-linear and $k_{\Psi,\mathrm{c}}$ which determines the position of the peak of $\mathcal{P}_\Psi(k,\eta_\mathrm{d})$. Here we need to note for completeness that $k_\mathrm{UV}$, $k_\mathrm{d}$ and $k_\mathrm{evap}$ are directly related to $M_\mathrm{PBH}$ and $\Omega_\mathrm{PBH,f}$ as~\cite{Domenech:2020ssp}
\beq\label{eq:k_evap_d_UV}
\begin{split}
&\frac{k_\mathrm{UV}}{k_\mathrm{f}} = \left(\frac{\Omega_\mathrm{PBH,f}}{\gamma}\right)^{1/3}\quad,\quad\frac{k_\mathrm{d}}{k_\mathrm{f}} = \sqrt{2} \Omega_\mathrm{PBH,f}\,,
\\ & \frac{k_\mathrm{evap}}{k_\mathrm{f}} = \left(\frac{3.8g_{*}\Omega_\mathrm{PBH,f}}{960C_\mathrm{G}\gamma}\right)^{1/3}\left(\frac{M_\mathrm{PBH}}{\Mp}\right)^{-2/3}\,, 
\end{split}
\eeq
where $k_\mathrm{f}$ reads
\beq\label{eq:k_f}
\frac{k_\mathrm{f}}{10^{20} \mathrm{Mpc}^{-1}} \simeq \left(\frac{3.8g_{*}}{960C_\mathrm{G}\gamma^{3/4}}\right)^{2} \left(\frac{M_\mathrm{PBH}}{10^4\mathrm{g}}\right)^{-5/6}\left(\frac{\Omega_\mathrm{PBH,f}}{10^{-7}}\right)^{-1/3}.
\eeq

As expected from \Eq{eq:P_Psi_eta_d}, $\mathcal{P}_\Psi(k,\eta_\mathrm{d})\propto k^{7/3}$ for $k<k_{\Psi,\mathrm{c}}$ and $\mathcal{P}_\Psi(k,\eta_\mathrm{d})\propto k^{-5/3}$ for  $k>k_{\Psi,\mathrm{c}}$. On the other hand, when evolved through the PBH-eMD era, $\mathcal{P}_\Psi(k)$ gets drastically enhanced especially on small scales due to the exponential growth of the sub-horizon transfer function $T_\Psi(k\eta, \epsilon)$ introduced in \Eq{eq:Psi_eMD_R_1+epsilon_full}. In order then to be well within the perturbative regime we define a critical scale $k_\mathrm{c}$ below which $\mathcal{P}_\Psi(k<k_\mathrm{c},\eta_\mathrm{evap})>1$. As a consequence, in the following we will consider scales $k_\mathrm{evap}<k<k_\mathrm{c}$.

%%%%%%%%%%%%%%%%%%%%%%%SECTION 5 %%%%%%%%%%%%%%%%%%%%%%%%%%%%%%%%%%%%%%%%%%%%%
\section{The scalar induced gravitational wave signal}\label{sec:SIGW}
With regard to the second-order induced tensor modes in $f(R)$ gravity, one can show that the $(\times)$ and $(+)$ polarisation modes obey the following equation of motion~\cite{Zhou:2024doz,Kugarajh:2025rbt}
\begin{eqnarray}
\label{eq:ehfR}
    h_{\mathbf{k}}^{\lambda,(2)''}\left(\eta \right)+\left(2\mathcal{H} +\frac{F^{(0)'}}{F^{(0)}} \right)h_{\mathbf{k}}^{\lambda,(2)'}\left(\eta \right) +k^2 h_{\mathbf{k}}^{\lambda,(2)}\left(\eta \right)
    =\frac{4\left(\mathcal{S}_{\mathbf{k}}^{\lambda,(2)} \left(\eta \right)+\sigma^{\lambda,(2)}_{\mathbf{k}}\left( \eta \right)\right)}{F^{(0)}}  \ ,
\end{eqnarray}
where $\lambda = (\times)\;\mathrm{or}\; (+)$ and the source terms $\mathcal{S}_{\mathbf{k}}^{\lambda,(2)} \left(\eta \right)$ and $\sigma^{\lambda,(2)}_{\mathbf{k}}\left( \eta \right)$ read as 
\begin{eqnarray}\label{eq:Ss}
\mathcal{S}_{\mathbf{k}}^{\lambda,(2)}(\eta)&&=\int\frac{d^3p}{(2\pi)^{3/2}}e^{\lambda, ij}(\mathbf{k})p_ip_j\left[\left(1+\frac{4}{3(1+w)}\right)\Phi^{(1)}_{\mathbf{k}-\mathbf{p}}(\eta) \Phi^{(1)}_{\mathbf{p}}(\eta)+2\Psi^{(1)}_{\mathbf{k}-\mathbf{p}}(\eta) \Phi^{(1)}_{\mathbf{p}}(\eta) \right. \nonumber\\
&&\left.+\frac{8}{3(1+w)\mathcal{H}}\Psi^{(1)'}_{\mathbf{k}-\mathbf{p}}(\eta) \Phi^{(1)}_{\mathbf{p}}(\eta)+\frac{4}{3(1+w)\mathcal{H}^2}\Psi^{(1)'}_{\mathbf{k}-\mathbf{p}}(\eta) \Psi^{(1)'}_{\mathbf{p}}(\eta) \right. \nonumber\\
&&\left.-\Psi^{(1)}_{\mathbf{k}-\mathbf{p}}(\eta) \Psi^{(1)}_{\mathbf{p}}(\eta) \right] 
\\ && =  \int\frac{\mathrm{d}^3 \nonumber \boldmathsymbol{p}}{(2\pi)^{3}}e^{\lambda,ij}(\boldmathsymbol{k})p_i p_j\,{S_{\bm p}} {S_{\bm {k-p}}}F_\mathcal{S}(\eta,p,|\bm{k-p}|),
\end{eqnarray}

\begin{eqnarray}\label{eq:Ssigma}
\sigma_{\mathbf{k}}^{\lambda,(2)}(\eta)&&=\int\frac{d^3p}{(2\pi)^{3/2}}e^{\lambda, ij}(\mathbf{k})p_ip_j\left(2\Psi^{(1)}_{\mathbf{k}-\mathbf{p}} (\eta) F^{(1)}_{\mathbf{p}}(\eta)+ \left(F^{(0)}-1 \right)\left(3\Psi^{(1)}_{\mathbf{k}-\mathbf{p}} (\eta) \Psi^{(1)}_{\mathbf{p}}(\eta) \right.\right. \nonumber\\
&&\left.\left.-\Phi^{(1)}_{\mathbf{k}-\mathbf{p}} (\eta) \Phi^{(1)}_{\mathbf{p}}(\eta) \right) \right) \nonumber
\\ &&  =  \int\frac{\mathrm{d}^3 \boldmathsymbol{p}}{(2\pi)^{3}}e^{\lambda,ij}(\boldmathsymbol{k})p_i p_j\,{\sigma_{\bm p}} {\sigma_{\bm {k-p}}}F_\mathcal{\sigma}(\eta,p,|\bm{k-p}|),
\end{eqnarray}
where the polarisation tensors $\e^{\lambda, ij}$ read as~\cite{Capozziello:2011et}
\beq
e^{(+),ij}(\boldmathsymbol{k}) = \frac{1}{\sqrt{2}}
\begin{pmatrix}
1 & 0 & 0\\
0 & -1 & 0 \\ 
0 & 0 & 0 
\end{pmatrix}, \quad
e^{(\times),ij}(\boldmathsymbol{k}) = \frac{1}{\sqrt{2}}
\begin{pmatrix}
0 & 1 & 0\\
1 & 0 & 0 \\ 
0 & 0 & 0
\end{pmatrix}.
\eeq
and   
\begin{align}
\begin{split}
F_\mathcal{S}(\eta,p,|\bm{k-p}|)& = \left(1+\frac{4}{3(1+w)}\right)T_\Phi(|\mathbf{k}-\mathbf{p}|\eta) T_\Phi(p\eta)+ 2T_\Psi(|\mathbf{k}-\mathbf{p}|\eta) T_\Phi(p\eta)
\\ &
+ \frac{8}{3(1+w)\mathcal{H}}T_\Psi(|\mathbf{k}-\mathbf{p}|\eta) T_\Phi(p\eta) + \frac{4}{3(1+w)\mathcal{H}^2}T'_\Psi(|\mathbf{k}-\mathbf{p}|\eta) T'_\Psi(p\eta) 
\\ &  -T_\Psi(|\mathbf{k}-\mathbf{p}|\eta) T_\Psi(p\eta), 
\end{split}
\\
\begin{split}
F_\mathcal{\sigma}(\eta,p,|\bm{k-p}|)  &=2T_{\Psi}(|\mathbf{k}-\mathbf{p}|\eta) T_\mathrm{sc}(p\eta)+ \left(F^{(0)}-1 \right)\Bigl[3T_{\Psi}(|\mathbf{k}-\mathbf{p}|\eta) T_{\Psi}(p\eta)  
\\ & -T_{\Phi}(|\mathbf{k}-\mathbf{p}|\eta)T_{\Phi}(p\eta) \Bigr].
\end{split}
\end{align}
$T_\mathrm{sc}$ stands for the transfer function of the scalaron first order perturbation defined on sub-horizon scales as~\cite{Zhou:2024doz}
\beq
\phi^{(1)}_{\mathrm{sc},\boldmathsymbol{k}}(\eta) = F^{(1)}_{s,\boldmathsymbol{k}}(\eta) \equiv T_s(k\eta)\Psi^{(1)}_{\mathbf{k}}(\eta_\mathrm{d}).
\eeq
See the discussion in \Sec{sec:scalaron_contribution} for more details.

We need to emphasize here that, in the case of $f(R)$ gravity, one is inevitably met with the presence of an anisotropic source term for the SIGW signal denoted as $\sigma_{\mathbf{k}}^{\lambda,(2)}(\eta)$ due to the fact that $\Psi^{(1)}$ and $\Phi^{(1)}$ are in principle different. At the end, the solution for the $(\times)$ and $(+)$ second-order tensor perturbations will read as 
\begin{eqnarray}
    h_{\mathbf{k}}^{\lambda,(2)}\left(\eta \right)=\frac{4}{a(\eta)}\int_0^\eta\mathrm{d}\bar{\eta}G_\boldmathsymbol{k}(\eta,\bar{\eta})
    a(\bar{\eta})\left[\frac{\mathcal{S}_{\mathbf{k}}^{\lambda,(2)}(\bar{\eta})+ \sigma_{\mathbf{k}}^{\lambda,(2)}(\bar{\eta})}{F^{(0)}(\bar{\eta})}\right] , \label{eq:sol1}
\end{eqnarray}
where $G_\boldmathsymbol{k}(\eta,\bar{\eta})$ is the Green's function, being the solution of the homogeneous equation
\beq\label{eq:Green_function_eq}
G^{\prime\prime}_\boldmathsymbol{k}(\eta,\bar{\eta}) +\frac{1}{F^{(0)}}\left(2\mathcal{H}F^{(0)}+F^{(0)'} \right)G^\prime_\boldmathsymbol{k}(\eta,\bar{\eta})
    +k^2G_\boldmathsymbol{k}(\eta,\bar{\eta}) = \delta\left(\eta - \bar{\eta}\right),
\eeq
with boundary conditions $\lim_{\eta\to \bar{\eta}}G^s_\boldmathsymbol{k}(\eta,\bar{\eta}) = 0$ and $ \lim_{\eta\to \bar{\eta}}G^{s,\prime}_\boldmathsymbol{k}(\eta,\bar{\eta})=1$. Finally, the second order tensor power spectrum will be recast as~\cite{Kohri:2018awv}
\begin{equation}\label{eq:PhPS}  
\overline{\mathcal{P}^{(2)}_{h,\lambda}(\eta,k)}=\frac{4}{{F^{(0)}}^2}\int_0^{\infty}dv\int_{|1-v|}^{1+v}du\left[\frac{4v^2-(1+v^2-u^2)^2}{4uv}\right]^2\tilde{\mathcal{P}}_S(kv)\tilde{\mathcal{P}}_S(ku)\overline{I^2}(x,u,v)~;
\end{equation}
\begin{equation}\label{eq:Idef}  I(x,u,v)=\int_{x_i}^{x}d\tilde{x}\,G(x,\tilde{x})f(\tilde{x},v,u)~, 
\end{equation}
where $x \equiv k\eta$, $ u \equiv |\bm{k-p}|/k $, $v = p/k$ and $f(\tilde{x},v,u)\equiv F_\mathcal{S}(\eta,vk,uk)+F_\mathcal{\sigma}(\eta,vk,uk)$. ${\tilde{\cal P}}_S(k)$ stands for the power spectrum of the PBH induced gravitational potentials $\Psi^{(1)}$ or $\Phi^{(1)}$ at the onset of the PBH-dominated era multiplied with the suppression factor $S^2_G(k)$. See \Eq{eq:P_Psi_eff} and \Eq{eq:P_Phi_eff} for the $\Psi\Psi$ and the $\Phi\Phi$ terms or the equivalent mixed terms $\Phi\Psi$, depending on which contribution in the source terms \eqref{eq:Ss} and \eqref{eq:Ssigma} dominate~\footnote{We need to note here that in the sub-horizon regime, where $\Psi^{(1)}$ and $\Phi^{(1)}$ are in principle different, $\Psi^{(1)}$ and $\Phi^{(1)}$ are related to each other according to \Eq{eq:Phi_vs_Phi}.} while $x_i$ denotes the initial time which we choose it to be the time of PBH formation $\eta_\mathrm{f}$.

%%%%%%%%%%%%%%%%%%%%%%%%%%%%%%%%%%%%%%%%%%%
\subsection{The dominant source of the scalar induced gravitational wave signal}

From \Eq{eq:Ss} and \Eq{eq:Ssigma}, we see that the second-order tensor modes $h_{\mathbf{k}}^{\lambda,(2)}\left(\eta \right)$ are sourced by the scalar gravitational potentials $\Psi^{(1)}$ and $\Phi^{(1)}$ and their derivatives $\Psi^{(1)'}$ and $\Phi^{(1)'}$. In \Eq{eq:Ssigma}, we have the appearance as well of the second derivative $\Psi^{(1)''}$ through the $F^{(1)}_\textbf{p}$ term. In GR, the dominant contribution was coming from the derivative terms $\mathcal{H}^{-2}\Psi^{(1)'}_{\mathbf{k}-\mathbf{p}}(\eta) \Psi^{(1)'}_{\mathbf{p}}(\eta) $ since the time derivative $\mathcal{H}^{-1}\Psi^{(1)'}$ was changing very suddenly from a zero value $\mathcal{H}^{-1}\Psi^{(1)'}=0$ during the PBH-eMD era  to $\mathcal{H}^{-1}\Psi^{(1)'}=(k\eta_\mathrm{evap})\Psi^{(1)}\gg \Psi^{(1)} $ right after PBH evaporation. In $f(R)$ gravity one needs to check if this condition holds as well. Let us then focus on the $R^{1+\epsilon}$ gravity since $R^2$ gravity, as we saw in \Sec{sec:scalar_perturbations}, behaves like in GR at the perturbative level. Doing so, from \Eq{eq:T_Psi_RD_FR}, one can show that at the end of PBH evaporation, i.e. at $\eta = \eta_\mathrm{evap}$, the ratio $\mathcal{H}^{-1}\Psi^{(1)'}/\Psi^{(1)}$ will read as
\begin{align}
    & \frac{\mathcal{H}^{-1}(\eta_\mathrm{evap})\Psi^{(1)'}(\eta_\mathrm{evap})}{\Psi^{(1)}(\eta_\mathrm{evap})}   \nonumber = \Biggl\{6k^4\eta^4_\mathrm{evap}\epsilon\left[254592 + k^4\eta^4_\mathrm{evap}\epsilon\left(816 + k^4\eta^4_\mathrm{evap}\epsilon\right)\right]\Biggr\}\\ & /\Biggl\{763776k^4\eta^4_\mathrm{evap}\epsilon + 1224k^8\eta^8_\mathrm{evap}\epsilon^2  + k^{12}\eta^{12}_\mathrm{evap}\epsilon^3 +164975616 %6873984 (24 +5\sqrt{6\epsilon}\;_1F_1\left(\frac{7}{4},\frac{7}{2},0\right)
    \Biggr\}.  
\end{align}

One then can show that for $x_\mathrm{evap}\equiv k\eta_\mathrm{evap}\gg 1$, $|\mathcal{H}^{-1}\Psi^{(1)'}/\Psi^{(1)}|=|\mathcal{H}^{-1}\Phi^{(1)'}/\Phi^{(1)}|>6>1$. We will consider then in the following only the derivative terms $\mathcal{H}^{-1}\Psi^{(1)'}$ and $\mathcal{H}^{-1}\Phi^{(1)'}$ which are at least $6^2= 36$ times larger than the terms $\Psi^{(1)}\Psi^{(1)}$, $\Psi^{(1)}\Phi^{(1)}$ and $\Psi^{(1)'}\Phi^{(1)}$ present in \Eq{eq:Ss} and \Eq{eq:Ssigma}.

Furthermore, one should consider 
as well the contribution $-\frac{6F^{(0)}_R}{a^2}\Psi^{(1)''}\Psi^{(1)}$ present in the anisotropic stress source term \eqref{eq:Ssigma} through the term $F^{(1)}_\mathbf{p}\Psi^{(1)}_\mathbf{k-p}$. Using thus \Eq{eq:T_Psi_RD_FR} one can show that
\beq\label{eq:Psi_prime_prime_o_Psi_prime}
\begin{split}
& \frac{\frac{6F^{(0)}_R}{a^2}\Psi^{(1)''}\Psi^{(1)}}{\mathcal{H}^{-2}\Psi^{(1)'}\Psi^{(1)'}}\Bigg\rvert_{\eta=\eta_\mathrm{evap}}  = -2\times 10^{-23}87^{\epsilon}\left(\frac{M_\mathrm{PBH}}{10^9\mathrm{g}}\right)^{3}\left(\frac{\Mp}{M_\mathrm{PBH}}\right)^{6\epsilon}\left(\frac{\Mp}{M_\mathrm{s}}\right)^{2\epsilon}\epsilon(1+\epsilon)\\ & \times  \Biggl[164975616 (2 + c^2_\mathrm{s}x^2_\mathrm{evap}) + 
       763776 x^4_\mathrm{evap} (18 + c^2_\mathrm{s}x^2_\mathrm{evap})\epsilon  + 
       1224 x^8_\mathrm{evap} (34 + c^2_\mathrm{s}x^2_\mathrm{evap})\epsilon^2 \\ &+ 
      x^{12}_\mathrm{evap} (50 + c^2_\mathrm{s}x^2_\mathrm{evap}) \epsilon^3 \Biggr] \Biggl[763776 x^4_\mathrm{evap} \epsilon + 
       1224x^8_\mathrm{evap}\epsilon^2 + x^{12}_\mathrm{evap}\epsilon^3 + 164975616  
       %6873984 \left[(24 + 5 \sqrt{6}\sqrt{\epsilon} \;_1F_1\left(\frac{7}{4},\frac{7}{2},0\right)\right]
       \Biggr] \\ & 
       /\Biggl\{ 144 x^8_\mathrm{evap}\epsilon^2 \left[254592 + 
       x^4_\mathrm{evap}\epsilon (816 + x^4_\mathrm{evap}\epsilon)\right]^2\Biggr\}.
\end{split}
\eeq
In the limit of $\epsilon\ll 1$ and $x_\mathrm{evap}\gg 1$, we obtain that 
\beq\label{eq:eq:Psi_prime_prime_o_Psi_prime_new}
\frac{\frac{6F^{(0)}_R}{a^2}\Psi^{(1)''}\Psi^{(1)}}{\mathcal{H}^{-2}\Psi^{(1)'}\Psi^{(1)'}}\Bigg\rvert_{\eta=\eta_\mathrm{evap}}\simeq 2\times 10^{-23}\left(\frac{M_\mathrm{PBH}}{10^9\mathrm{g}}\right)^{3}\epsilon\Bigl[\frac{c^2_\mathrm{s}x^2_\mathrm{evap}}{144} + \frac{25}{72}\Bigr]
\eeq
%\beq\label{eq:eq:Psi_prime_prime_o_Psi_prime_new}
%\frac{\frac{6F^{(0)}_R}{a^2}\Psi^{(1)''}\Psi^{(1)}}{\mathcal{H}^{-2}\Psi^{(1)'}\Psi^{(1)'}}\Bigg\rvert_{\eta=\eta_\mathrm{evap}}\simeq -\frac{2c^2_\mathrm{s}(6+7\epsilon)}{k^2\eta^2_\mathrm{evap}} \ll 1
%\eeq
From \Eq{eq:eq:Psi_prime_prime_o_Psi_prime_new}, one can clearly see that since $M_\mathrm{PBH}<10^9\mathrm{g}$, $\Biggl|\frac{\frac{6F^{(0)}_R}{a^2}\Psi^{(1)''}\Psi^{(1)}}{\mathcal{H}^{-2}\Psi^{(1)'}\Psi^{(1)'}}\Bigg\rvert_{\eta=\eta_\mathrm{evap}}\ll 1$. One then can safely neglect the anisotropic source term \Eq{eq:Ssigma} in the following analysis becoming subdominant compared to the \eqref{eq:Ss} source term, something which was also noted as well in~\cite{Kugarajh:2025rbt} but for the case of $R^2$ gravity.

%As a consequence, their contribution to the GW amplitude being quadratically dependent on the gravitational potentials will be at least one order of magnitude higher compared to the contribution to the other terms. In this way, the anisotropic source term \Eq{eq:Ssigma} becomes subdominant compared to the \eqref{eq:Ss} source term, something which was also noted as well in~\cite{Kugarajh:2025rbt}.

Before going to the computation of the SIGW signal for the $(\times)$ and $(+)$ polarisations it is important to stress here as well that the kernel function $I(x,u,v)$ present in \Eq{eq:PhPS} have three contributions and can be written as
\beq
I(x,u,v) = I_\mathrm{eRD}(x,u,v) +  I_\mathrm{eMD}(x,u,v) +  I_\mathrm{lRD}(x,u,v).
\eeq
From the three terms in the kernel function, the last one, related to the late RD (lRD) era, is the dominant one due to the enhanced amplitude of the oscillations of $\Psi$ and $\Phi$ during the lRD era~\cite{Inomata:2019ivs}. Moreover, since we consider nearly monochromatic PBH mass functions, we are met inevitably with a sudden transition from the eMD era to the lRD era, leading to no correlation between perturbation modes produced in different cosmic epochs~\cite{Inomata:2019ivs}. Thus, oscillation average cross-product terms such as $\overline{I_\mathrm{eMD}I_\mathrm{lRD}}$, are not considered in the following~\cite{Domenech:2020ssp}.
%%%%%%%%%%%%%%%%%%%%%%%%%%%%%
\subsection{The kernel function \texorpdfstring{$I(u,v,x)$}{I(u,v,x)}}
Considering modes being deep in the sub-horizon regime at the end of the PBH-eMD era, i.e. $x_\mathrm{evap}=k\eta_{\rm evap}\gg 1$, one can recast $I(u,v,x)$ in the lRD era as 
\begin{equation}\label{eq:IlRD}
    I_{\rm lRD}(x,u,v,x_{\rm evap})\simeq\frac{1}{2}uv\int_{x_{\rm evap}/2}^{x}\mathrm{d}\tilde{\bar{x}}\,\tilde{\bar{x}}^2\,G(x,\tilde{\bar{x}})\frac{\mathrm{d} T^\mathrm{R^{1+\epsilon},lRD}_{\Psi}(u\tilde{\bar{x}})}{\mathrm{d}(u\tilde{\bar{x}})}\frac{\mathrm{d} T^\mathrm{R^{1+\epsilon},lRD}_{\Psi}(v\tilde{\bar{x}})}{\mathrm{d}(v\tilde{\bar{x}})} \,,
\end{equation}
where $G( x,{\tilde {\bar x}})=\frac{{\tilde {\bar x}}}{ x}\sin(x- {\tilde {\bar x}})$ is the Green's function during the lRD era~\cite{Papanikolaou:2021uhe}. From \Eq{eq:T_Psi_RD_FR} one gets that for the scales of our interest, namely for $x_\mathrm{evap}\gg 1$, 
\begin{align}\label{eq:tprime}
\frac{d}{d(v \tilde{\bar x})}T^\mathrm{R^{1+\epsilon},lRD}_{\Psi}(v\tilde{\bar x})& \approx T_{\Psi,\mathrm{eMD}}(x_\mathrm{evap},\epsilon)\left(\frac{x_{\rm evap}}{2\tilde{\bar x}}\right)^2 \Biggl[-c^2_\mathrm{s}\sin\left[c_s v\left(\tilde{\bar x}-\frac{x_{\rm evap}}{2}\right)\right] \\ & + \frac{x^3_{\rm evap}}{2^4 3^3}\cos\left[c_s v\left(\tilde{\bar x}-\frac{x_{\rm evap}}{2}\right)\right]\epsilon \Biggr]\nonumber.
\end{align}
Interestingly, for $\epsilon\rightarrow 0$, the second term vanishes and one recovers the GR result, namely Eq. (C2) of~\cite{Domenech:2020ssp}.
Since we are interested in computed the induced GW signal at late times we can safely take $x\to\infty$. Doing so, and making a Taylor expansion for $\epsilon\ll 1$, \Eq{eq:IlRD} will be recast as 
\beq\label{eq:kernelRD2}
\begin{split}
I_\mathrm{lRD}(x,u,v,x_{\rm evap})\approx & \frac{uv}{32x} x_{\rm evap}^4T_{\Psi,\mathrm{eMD}}(x_\mathrm{evap},\epsilon)\\ & \times \Biggl[c^2_\mathrm{s}\int_{0}^{\infty} \frac{dx_2}{x_2+x_{\rm evap}/2}\sin(x_1-x_2)\sin(vc_s x_2)\sin(uc_s x_2) \\ & -\frac{x^6_\mathrm{evap}}{2^83^6}\epsilon^2\int_{0}^{\infty} \frac{dx_2}{x_2+x_{\rm evap}/2}\sin(x_1-x_2)\cos(vc_s x_2)\cos(uc_s x_2) \Biggr] ,
\end{split}
\eeq
where $x_1\equiv x-x_{\rm evap}/2$ and $x_2\equiv\tilde{\bar x}-x_{\rm evap}/2$. In ~\eqref{eq:kernelRD2} we have ignored the integrals with cross terms $\sin(uc_\mathrm{s}x_2)\cos(uc_\mathrm{s}x_2)$ since they give a zero contribution. The integral~\eqref{eq:kernelRD2} diverges when the frequency of the faster oscillating mode is equal to the sum of frequencies of the slower oscillating modes, namely when $u+v=c_s^{-1}$~\cite{Domenech:2021ztg}. One then at the end can compute analytically the integral~\eqref{eq:kernelRD2} and select the resonant  contributions where $u+v=c_s^{-1}$. Finally, we get that $\overline{I^2_{\rm lRD,{\rm res}}}(x,u,v,x_\mathrm{evap})$ reads as ~\cite{Inomata:2019ivs,Domenech:2020ssp}
\begin{align}\label{eq:irdapp}
\overline{I^2_{\rm lRD,{\rm res}}}(x,u+v\sim c_s^{-1},x_\mathrm{evap})\simeq &  \frac{u^2v^2x^8_\mathrm{evap}}{2^{15}x^2}\Biggl[c^4_\mathrm{s} + \frac{x^{12}_\mathrm{evap}\epsilon^4}{2^{16}3^{12}}-\frac{c^2_\mathrm{s}x^{6}_\mathrm{evap}\epsilon^2}{2^83^6}
\Biggr]\\ & \times T^2_{\Psi,\mathrm{eMD}}(x_\mathrm{evap},\epsilon) {\rm Ci}^2(|1-(u+v)c_s|x_{\rm evap}/2) \nonumber.
\end{align}

The second dominant contribution to $I_{\rm lRD}(x,u,v,x_\mathrm{evap})$ comes when $u\sim v\gg1$, i.e. when the integrand of \Eq{eq:PhPS} diverges. At the end, following the analysis present in the Appendix C of~\cite{Domenech:2020ssp} the large $v,u$ contribution $\overline{I^2_{\rm lRD,{\rm LV}}}(x,u,v,x_\mathrm{evap})$ can be recast as
\begin{align}\label{eq:irdapp2}
\overline{I^2_{\rm lRD,{\rm LV}}}(x, u\sim v\gg 1,x_\mathrm{evap})\simeq &\frac{u^2v^2}{2^{13}x^2} x_{\rm evap}^{8} \Biggl[c^4_\mathrm{s} + \frac{x^{12}_\mathrm{evap}\epsilon^4}{2^{16}3^{12}}-\frac{c^2_\mathrm{s}x^{6}_\mathrm{evap}\epsilon^2}{2^83^6}
\Biggr] \\ & \times T^2_{\Psi,\mathrm{eMD}}(x_\mathrm{evap},\epsilon)\Big[{\rm Ci}^2(x_{\rm evap}/2)+\big({\rm Si}(x_{\rm evap}/2)-\pi/2\big)^2\Big]\nonumber .
\end{align}
We need to note here that in both the resonant and $\mathrm{LV}$ kernels \eqref{eq:irdapp} and \eqref{eq:irdapp2} we accounted for the effect of $f(R)$ gravity whereas for $\epsilon\rightarrow 0$, one recovers the GR expressions for $\overline{I^2_{\rm lRD,{\rm res}}}(x,u+v\sim c_s^{-1},x_\mathrm{evap})$ and $\overline{I^2_{\rm lRD,{\rm LV}}}(x, u\sim v\gg 1,x_\mathrm{evap})$, namely Eq (C5) and (C6) of~\cite{Domenech:2020ssp}.
 
%%%%%%%%%%%%%%%%%%%%%%%%%%%%%%%%%%%%%%%%%%%%%
\subsection{The scalar induced gravitational wave spectrum}
Having derived above the second order induced tensor power spectrum and the dominant contributions to the kernel function $I_{\rm lRD}(x,u,v)$ one can compute the GW spectral abundance $\Omega_\mathrm{GW}(\eta,k)$ at a time $\eta$. Focusing here on the $(\times)$ and the $(+)$ polarisation modes, the GW spectral abundance on sub-horizon scales will read as ~\cite{Maggiore:1999vm}
\beq\label{Omega_GW_GR}
\Omega_\mathrm{GW}(\eta,k) = 
\frac{1}{24}\left(\frac{k}{\calH(\eta)}\right)^{2}\overline{\mathcal{P}^{(2)}_{h,\lambda}(\eta,k)}
\eeq
where $\mathcal{P}^{(2)}_{h,\lambda}(\eta,k)$ is given by \Eq{eq:PhPS} and the bar denotes an oscillation average. Accounting then for the fact that the 
radiation energy 
density reads as $\rho_r = 
\frac{\pi^2}{30}g_{*\mathrm{\rho}}T_\mathrm{r}^4$ and that the temperature of the radiation thermal bath $T_\mathrm{r}$ scales as $T_\mathrm{r}\propto 
g^{-1/3}_{*\mathrm{S}}a^{-1}$, one gets that $\Omega_\mathrm{GW}(\eta,k)$ at 
our present epoch will be recast as~\cite{Espinosa:2018eve}
\beq\label{Omega_GW_RD_0}
\Omega_\mathrm{GW}(\eta_0,k) = 
\Omega^{(0)}_\mathrm{r}\frac{g_{*\mathrm{\rho},\mathrm{*}}}{g_{*\mathrm{\rho},0}}
\left(\frac{g_{*\mathrm{S},\mathrm{0}}}{g_{*\mathrm{S},\mathrm{*}}}\right)^{4/3}
\Omega_\mathrm{GW}(\eta_\mathrm{*},k),
\eeq
where $g_{*\mathrm{\rho}}$ and $g_{*\mathrm{S}}$ denote the energy and 
entropy relativistic degrees of freedom and $\eta_*$ stands for a reference conformal time, after which GWs propagate freely. In particular, $\eta_*$ should be deep enough in the late RD era in order for the GW source to decay enough so as that we can treat consistently the induced tensor modes as free propagating GWs. In our case, as one can see from \Eq{eq:PhPS} and \Eq{Omega_GW_GR}, the time dependence of $\Omega_\mathrm{GW}$ in the late RD era enters in three places, namely in $\overline{I^2_\mathrm{lRD}(x,u,v)}$, in ${F^{(0)}}^2$ and in $(k/\mathcal{H})^2$. However, since during the late RD era ${F^{(0)}}^2\simeq 1$ - since one revovers GR in the late RD for viable power-law $f(R)$ gravity models~\cite{Odintsov:2025eiv,Fazzari:2025nfr}, $(k/\mathcal{H})^2\propto (k\eta)^2\propto x^2$ and $\overline{I^2_{\rm lRD}(x,u,v)}\propto x^{-2}$ as we can see from \Eq{eq:irdapp} and \Eq{eq:irdapp2}, one can basically choose safely whatever value for $\eta_*$ greater than $\eta_\mathrm{evap}$. This result regarding the choice of $\eta_\mathrm{evap}$ was also found in~\cite{Domenech:2020ssp}.

\subsubsection{The resonant contribution \texorpdfstring{$u+v= c^{-1}_\mathrm{s}$}{u+v= c^{-1}_s}}

Focusing on the resonant contribution $u+v= c^{-1}_\mathrm{s}$ one can insert \Eq{eq:PhPS} into \Eq{Omega_GW_GR} with the averaged kernel function given by \Eq{eq:irdapp}. Finally, $\Omega^\mathrm{res}_\mathrm{GW}(\eta_*,k)$ will read as
\beq\label{eq:Omega_GW_res_full}
\begin{split}
\Omega^\mathrm{res}_\mathrm{GW}(\eta_*,k) & = c_g\Omega^{(0)}_\mathrm{r}\epsilon^2\frac{1}{2^83^3}\left(\frac{3}{2}\right)^{\frac{2}{3}}\left(\frac{k}{k_\mathrm{evap}}\right)^{8}\left(\frac{k}{k_\mathrm{UV}}\right)^{14/3}\left(\frac{k_\mathrm{evap}}{k_\mathrm{UV}}\right)^{4/3} \\ & \times 
\Biggl[c^4_\mathrm{s} + \frac{\epsilon^4}{2^{4}3^{12}}\left(\frac{k}{k_\mathrm{evap}}\right)^{12}-\frac{c^2_\mathrm{s}\epsilon^2}{2^23^6}\left(\frac{k}{k_\mathrm{evap}}\right)^{6}
\Biggr]
\\ & \int_{k_\mathrm{evap}/k}^{k_\mathrm{c}/k}\mathrm{d}v\int_{\max[|1-v|,k_\mathrm{evap}/k]}^{\min[1+v,k_\mathrm{c}/k]}\mathrm{d}u\left(\frac{4v^2-(1+v^2-u^2)^2}{4uv}\right)^2\left(uv\right)^{13/3} 
\\ & \times \mathrm{Ci}^2\left(|1-(u+v)c_\mathrm{s}|\frac{k}{k_\mathrm{evap}}\right) \\ & \times T^2_{\Psi,\mathrm{eMD}}(kv\eta_\mathrm{evap},\epsilon)T^2_{\Psi,\mathrm{eMD}}(ku\eta_\mathrm{evap},\epsilon)T^2_{\Psi,\mathrm{iso}}(kv\eta_\mathrm{d},\epsilon)T^2_{\Psi,\mathrm{iso}}(ku\eta_\mathrm{d},\epsilon),
\end{split}
\eeq
where $T_{\Psi,\mathrm{eMD}}(k\eta,\epsilon)$ is given by \Eq{eq:Psi_eMD_R_1+epsilon} and $T_{\Psi,\mathrm{iso}}(k\eta,\epsilon)$ is defined as
\beq
T_{\Psi,\mathrm{iso}}(k\eta,\epsilon) 
\equiv\left[5+\frac{2}{3}\left(\frac{k}
{\mathcal{H}}\right)^2F^{(0)}\left(\frac{1+ 3 \frac{k^2}{a^2}\frac{F^{(0)}_{R}}{F^{(0)}}}{1+ 2 \frac{k^2}{a^2} \frac{F^{(0)}_{R}}{F^{(0)}}}\right)\right]^{-1}.
\eeq
In order to simplify the above integral, one can make a variable transformation given by
\begin{align}
y & \equiv  ((u+v)c_\mathrm{s}-1)x_\mathrm{evap}/2,\\ 
s & \equiv u - v,
\end{align}
whose Jacobian reads as $|J| = \frac{1}{c_\mathrm{s}x_\mathrm{evap}}$,
where $x = k\eta_\mathrm{evap}=2k/k_\mathrm{evap}$. One then can use the fact that $\int_{-\infty}^{\infty}\mathrm{Ci}^2(|y|)\mathrm{d}y = 1$ to get rid of one integral in \Eq{eq:Omega_GW_res_full}. At the end, \Eq{eq:Omega_GW_res_full} will be recast as
\beq\label{eq:Omega_GW_res_approx}
\begin{split}
\Omega^\mathrm{res}_\mathrm{GW}(\eta_*,k) & = c_g\Omega^{(0)}_\mathrm{r}\epsilon^2\frac{(1 - c^2_\mathrm{s})^2}{2^{16}3^3c_\mathrm{s}^8}\left(\frac{3}{4}\right)^{\frac{2}{3}}\left(\frac{k}{k_\mathrm{evap}}\right)^{8}\left(\frac{k}{k_\mathrm{UV}}\right)^{14/3}\left(\frac{k_\mathrm{evap}}{k_\mathrm{UV}}\right)^{4/3} \\ & \times 
\Biggl[c^4_\mathrm{s} + \frac{\epsilon^4}{2^{4}3^{12}}\left(\frac{k}{k_\mathrm{evap}}\right)^{12}-\frac{c^2_\mathrm{s}\epsilon^2}{2^23^6}\left(\frac{k}{k_\mathrm{evap}}\right)^{6}
\Biggr]\\ & \int_{-s_0(k)}^{s_0(k)}\mathrm{d}s (1-s^2)^2(1-c^2_\mathrm{s}s^2)^{7/3} \\ & \times T^2_{\Psi,\mathrm{eMD}}\left[k\left(\frac{1-c_\mathrm{s}s}{2c_\mathrm{s}}\right)\eta_\mathrm{evap},\epsilon\right]T^2_{\Psi,\mathrm{eMD}}\left[k\left(\frac{1+c_\mathrm{s}s}{2c_\mathrm{s}}\eta_\mathrm{evap}\right),\epsilon\right] \\ & \times T^2_{\Psi,\mathrm{iso}}\left[k\left(\frac{1-c_\mathrm{s}s}{2c_\mathrm{s}}\right)\eta_\mathrm{d}\right]T^2_{\Psi,\mathrm{iso}}\left[k\left(\frac{1-c_\mathrm{s}s}{2c_\mathrm{s}}\right)\eta_\mathrm{d}\right],
\end{split}
\eeq
where $s_0(k)$ is given by
\begin{align}\label{eq:s0}
s_0(k)=\left\{
\begin{aligned}
&1\qquad & \tfrac{k_{\rm c}}{k}\geq\tfrac{1+c_s^{-1}}{2}\\
&2\tfrac{k_{\rm c}}{k}-c_s^{-1}\qquad & \tfrac{1+c_s^{-1}}{2}\geq\tfrac{k_{\rm c}}{k}\geq\tfrac{c_s^{-1}}{2}\\
&0\qquad &\tfrac{c_s^{-1}}{2}\geq\tfrac{k_{\rm c}}{k}
\end{aligned}
\right.\,.
\end{align} 
%%%%%%%%%%%%%%%%%%%%%%%%%%%%%%%%%%%%%%%%

\subsubsection{The large \texorpdfstring{$v$}{v} contribution }
Concerning now the large $v$ (LV) contributions, one can straightforwardly write $\Omega^\mathrm{LV}_\mathrm{GW}(\eta_\mathrm{evap},k)$ as
\beq\label{eq:Omega_GW_LV_full}
\begin{split}
\Omega^\mathrm{LV}_\mathrm{GW}(\eta_*,k) & = c_g\Omega^{(0)}_\mathrm{r}\epsilon^2\frac{1}{2^63^3}\left(\frac{3}{2}\right)^{\frac{2}{3}}\left(\frac{k}{k_\mathrm{evap}}\right)^{8}\left(\frac{k}{k_\mathrm{UV}}\right)^{14/3}\left(\frac{k_\mathrm{evap}}{k_\mathrm{UV}}\right)^{4/3} \\ & \times 
\Biggl[c^4_\mathrm{s} + \frac{\epsilon^4}{2^{4}3^{12}}\left(\frac{k}{k_\mathrm{evap}}\right)^{12}-\frac{c^2_\mathrm{s}\epsilon^2}{2^23^6}\left(\frac{k}{k_\mathrm{evap}}\right)^{6}
\Biggr]\\ & \int_{k_\mathrm{evap}/k}^{k_\mathrm{c}/k}\mathrm{d}v\int_{\max[|1-v|,k_\mathrm{evap}/k]}^{\min[1+v,k_\mathrm{c}/k]}\mathrm{d}u\left(\frac{4v^2-(1+v^2-u^2)^2}{4uv}\right)^2\left(uv\right)^{13/3} 
\\ & \times \left[\mathrm{Ci}^2\left(\frac{k}{k_\mathrm{evap}}\right)^2 +\left(\mathrm{Si}\left(\frac{k}{k_\mathrm{evap}}\right) -\frac{\pi}{2}\right)^2\right] \\ & \times T^2_{\Psi,\mathrm{eMD}}(kv\eta_\mathrm{evap},\epsilon)T^2_{\Psi,\mathrm{eMD}}(ku\eta_\mathrm{evap},\epsilon)T^2_{\Psi,\mathrm{iso}}(kv\eta_\mathrm{d},\epsilon)T^2_{\Psi,\mathrm{iso}}(ku\eta_\mathrm{d},\epsilon).
\end{split}
\eeq

Making then again a variable transformation given by
\begin{align}
t & \equiv  u+v-1,\\ 
s & \equiv u - v,
\end{align}
with a Jacobian $|J| = \frac{1}{2}$, \Eq{eq:Omega_GW_LV_full} takes the following form:
\beq\label{eq:Omega_GW_LV_approx}
\begin{split}
\Omega^\mathrm{LV}_\mathrm{GW}(\eta_*,k) & = c_g\Omega^{(0)}_\mathrm{r}\epsilon^2\frac{1}{2^{15}3^3}\left(\frac{3}{4}\right)^{\frac{2}{3}}\left(\frac{k}{k_\mathrm{evap}}\right)^{8}\left(\frac{k}{k_\mathrm{UV}}\right)^{14/3}\left(\frac{k_\mathrm{evap}}{k_\mathrm{UV}}\right)^{4/3} \\ & \times 
\Biggl[c^4_\mathrm{s} + \frac{\epsilon^4}{2^{4}3^{12}}\left(\frac{k}{k_\mathrm{evap}}\right)^{12}-\frac{c^2_\mathrm{s}\epsilon^2}{2^23^6}\left(\frac{k}{k_\mathrm{evap}}\right)^{6}
\Biggr] \\ & \int_{2k_\mathrm{evap}/k - 1}^{2k_\mathrm{c}/k - 1}\mathrm{d}t t^2(1+t)^{14/3}(2+t)^2 \\ & \times T^4_{\Psi,\mathrm{eMD}}\left[k\left(\frac{1+t}{2}\right)\eta_\mathrm{evap},\epsilon\right]T^4_{\Psi,\mathrm{iso}}\left[k\left(\frac{1+t}{2}\right)\eta_\mathrm{d}\right],
\end{split}
\eeq
where we have set $s= u - v\simeq 0$, getting rid of one integral in \Eq{eq:Omega_GW_LV_full}.
%%%%%%%%%%%%%%%%%%%%%%%%%%%%%%%%%%%%%%%%%%%%%%
\subsubsection{The scalaron contribution}\label{sec:scalaron_contribution}
Let us then study in this section the contribution of the scalaron massive mode introduced in \Sec{sec:f_R_gravity_basics} at the level of GWs. As it was already found in~\cite{Papanikolaou:2021uhe} the contribution of the scalaron polarization mode on the second-order GWs induced by PBH energy density perturbations is negligible compared to the $(+)$ and $(\times)$ polarization contributions. However, in this section, we will study how the first-order scalaron perturbations can source first-order GWs following closely the analysis of~\cite{Zhou:2024doz}. Doing so, one should perturb \Eq{eq:scalaron_background} at first-order getting at the end that the first-order scalaron perturbation $\phi_\mathrm{sc}^{(1)}$ obeys the same wave equation as the second-order scalar induced GWs but with a different source term. In particular, $\phi_\mathrm{sc}^{(1)}$ obeys the following wave equation
\cite{Katsuragawa:2019uto}:
\begin{eqnarray}\label{eq:phis1}
\phi_\mathrm{sc}^{(1)''}+2\mathcal{H}\phi_\mathrm{sc}^{(1)'}-\left(\Delta+m^2_\mathrm{sc}  \right)\phi_\mathrm{sc}^{(1)}=-\square^{(1)} \phi_\mathrm{sc}^{(0)} +\frac{\kappa}{3} T^{M,(1)},
\end{eqnarray}
where $T^{M}$ is the trace of the matter stress-energy tensor. Given thus \Eq{eq:phis1}, one can consider the scalaron field $\phi_\mathrm{sc}$ as an additional massive polarisation contribution for the GW signal with source term: $\mathcal{S}^{(1)}_\mathrm{sc}=-\square^{(1)} \phi_\mathrm{sc}^{(0)} +\frac{\kappa}{3} T^{M,(1)}$. One then instead of solving numerically or through the Green's function formalism \Eq{eq:phis1}, like in the case of SIGWs, they can express the perturbed scalaron field $\phi_\mathrm{sc}^{(1)}$ in terms of the first-order Bardeen potentials $\Phi^{(1)}$ and $\Psi^{(1)}$ through its formal definition as $\phi_\mathrm{sc}\equiv \frac{\mathrm{d}f(R)}{\mathrm{d}R}$. More specifically, we can recast $\phi_{s}^{(1)}$ as 
\beq\label{eq:phi_s_1}
\phi_{\mathrm{sc},\mathbf{k}}^{(1)}(\eta) = F^{(1)}_{\mathbf{k}}(\eta) = F^{(0)}_R(\eta) R^{(1)}_{\mathbf{k}}(\eta), 
\eeq
where $ R^{(1)}_{\mathbf{k}}$ is the first order perturbed Ricci scalar, which in terms of $\Phi^{(1)}$ and $\Psi^{(1)}$ can be written as~\cite{Tsujikawa:2007gd}:
\beq\label{eq:Ricci_scalar_perturbation}
\begin{split}
R^{(1)}_{\mathbf{k}}(\eta) &  = -\frac{2}{a^2}\Bigl[6(\mathcal{H}^\prime +\mathcal{H}^2)\Phi^{(1)}_{\mathbf{k}}(\eta) + 3\mathcal{H}\left(\frac{\mathrm{d}\Phi^{(1)}_{\mathbf{k}}(\eta)}{\mathrm{d}\eta} + 4\frac{\mathrm{d}\Psi^{(1)}_{\mathbf{k}}(\eta)}{\mathrm{d}\eta}\right) + 3\frac{\mathrm{d}^2\Psi^{(1)}_{\mathbf{k}}(\eta)}{\mathrm{d}\eta^2} \\ & - k^2\Phi^{(1)}_{\mathbf{k}}(\eta) + 2k^2\Psi^{(1)}_{\mathbf{k}}(\eta)\Bigr].
\end{split}
\eeq

Working then on sub-horizon scales, which are the scales of interest for us since we consider scales within the cosmological horizon at PBH evaporation time, i.e. $k>k_\mathrm{evap}$, one can neglect the time derivative terms in \Eq{eq:Ricci_scalar_perturbation} and write $R^{(1)}_{\mathbf{k}}(\eta)$ as 

\beq
R^{(1)}_{\mathbf{k}}(\eta) \simeq k^2(2\Psi^{(1)}_{\mathbf{k}}(\eta) - \Phi^{(1)}_{\mathbf{k}}(\eta)) = -\frac{2}{a^2}\frac{\Psi^{(1)}_{\mathbf{k}}(\eta)}{2\frac{F^{(0)}_R}{F^{(0)}}\frac{k^2}{a^2}+1},
\eeq
where in the last equality we used \Eq{eq:Phi_vs_Phi}.

One then can define a transfer function for the scalaron on sub-horizon scales as follows: 
\beq
\phi^{(1)}_{\mathrm{sc},\boldmathsymbol{k}}(\eta) \equiv T_\mathrm{sc}(k\eta)\Psi^{(1)}_{\mathbf{k}}(\eta_\mathrm{d}),
\eeq
where $T_\mathrm{sc}$ can be written in terms of $T_\Psi$ as 
\beq
T_\mathrm{sc}(k\eta) = -\frac{2F^{(0)}_R}{a^2}\frac{T_\Psi(k\eta)}{2\frac{F^{(0)}_R}{F^{(0)}}\frac{k^2}{a^2}+1}.
\eeq
Followingly, the first order tensor power spectrum sourced by $\phi_{\mathrm{sc},\mathbf{k}}^{(1)}$ will be recast as
\beq
\mathcal{P}^{(1)}_{h,\mathrm{sc}}(\eta,k) = \frac{k^3|\phi^{(1)}_{s,\boldmathsymbol{k}}(\eta)|^2}{2\pi^2} = T^2_\mathrm{sc}(k\eta) \mathcal{P}_\Psi(k,\eta_\mathrm{d}),
\eeq
where $\mathcal{P}_\Psi(k,\eta_\mathrm{d})$ is given by \Eq{eq:P_Psi_eta_d} and the associated to it spectral abundance will read as 
\beq\label{Omega_GW_scalaron}
\Omega^\mathrm{sc}_\mathrm{GW}(\eta,k) = 
\frac{1}{6}\left(\frac{k}{\calH(\eta)}\right)^{2} \mathcal{P}^{(1)}_{h,\mathrm{sc}}(\eta,k).
\eeq
After some algebra, one can show that $\Omega^\mathrm{sc}_\mathrm{GW}$ at our present epoch will be given by
\beq\label{eq:Omega_GW_scalaron_approx}
\Omega^\mathrm{sc}_\mathrm{GW}(\eta_0,k) = \frac{c_g\Omega^{(0)}_r}{6}\left(\frac{k}{k_\mathrm{evap}}\right)^2\left[\frac{2F^{(0)}_{R,\mathrm{evap}}}{a^2_\mathrm{evap}}\frac{T_\Psi(k\eta_\mathrm{evap},\epsilon)}{2\frac{F^{(0)}_{R,\mathrm{evap}}}{F^{(0)}_\mathrm{evap}}\frac{k^2}{a^2_\mathrm{evap}}+1}\right]^2\mathcal{P}_\Psi(k,\eta_\mathrm{d}).
\eeq

%%%%%%%%%%%%%%%%%%%%%%%%%%%%%%%%%%%%%%%%%%%%%%%%%%
\section{Results}\label{sec:results}
\begin{figure}
\begin{center}
\includegraphics[width = 0.6\textwidth]{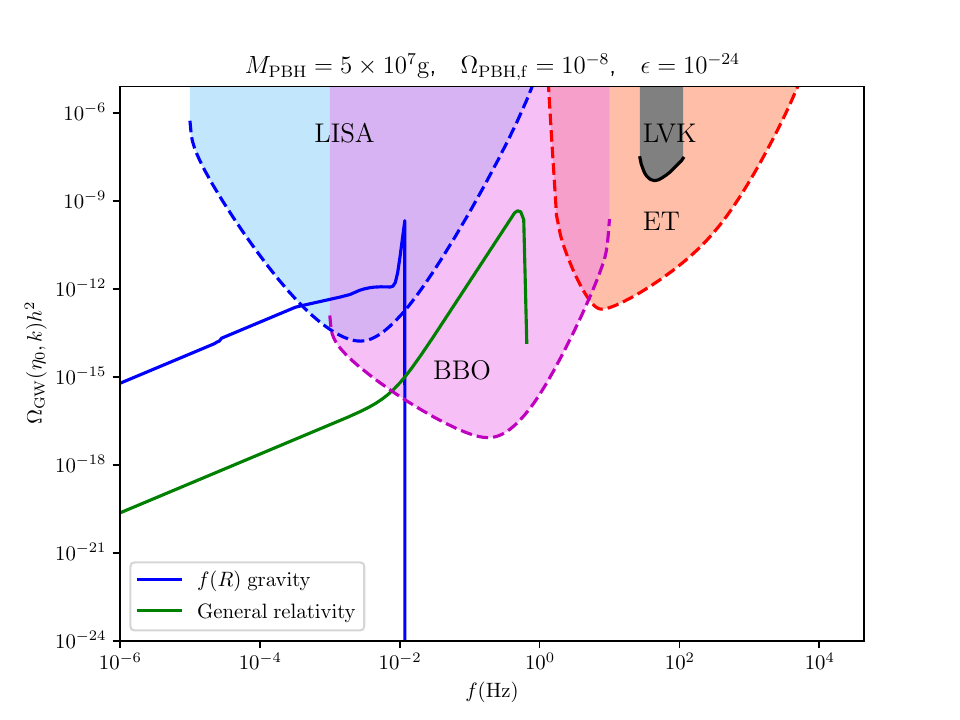}
\caption{{ The scalar-induced gravitational wave signal in $R^{1+\epsilon}$ gravity for $M_\mathrm{PBH} = 5\times 10^7\mathrm{g}$, $\Omega_\mathrm{PBH,f} = 10^{-8}$ and $\epsilon = 10^{-24}$ (blue curve) and the respective gravitational-wave signal in GR (green curve).}}
\label{fig:Omega_GW_FR_vs_GR}
\end{center}
\end{figure}

Having derived above the scalar-induced GW signal in $R^{1+\epsilon}$ gravity as well the dominant contribution to it, namely the resonant one \eqref{eq:Omega_GW_res_approx}, the large $v$ (LV) one \eqref{eq:Omega_GW_LV_approx} and the scalaron one \eqref{eq:Omega_GW_scalaron_approx} let us now see how the SIGW signal behaves. In \Fig{fig:Omega_GW_FR_vs_GR} we depict the SIGW signal in $R^{1+\epsilon}$ gravity for $M_\mathrm{PBH} = 5\times 10^7\mathrm{g}$, $\Omega_\mathrm{PBH,f} = 10^{-8}$ and $\epsilon =  10^{-24}$ (blue curve) and the respective gravitational-wave signal in GR (green curve). As one can see, the $R^{1+\epsilon}$ gravity SIGW signal peaks at a frequency $f_\mathrm{peak} \equiv \frac{k_\mathrm{c}}{2\pi}$ smaller compared to the GR one due to the fact that scalar perturbations grow exponentially within the PBH-eMD era, imposing thus a non-linear cut-off $k^{-1}_\mathrm{c}$ larger than the $k^{-1}_\mathrm{UV}$ GR cut-off. In addition, we also note the more peaky behaviour of the GW spectrum around $k_\mathrm{c}$ within $R^{1+\epsilon}$ gravity  compared to the GR GW spectrum behaviour around $k_\mathrm{UV}$. In the infrared (IR) regions both $f(R)$ and GR GW spectra scale linearly with the frequency.

\begin{figure}
\begin{center}
\includegraphics[width = 0.49\textwidth]{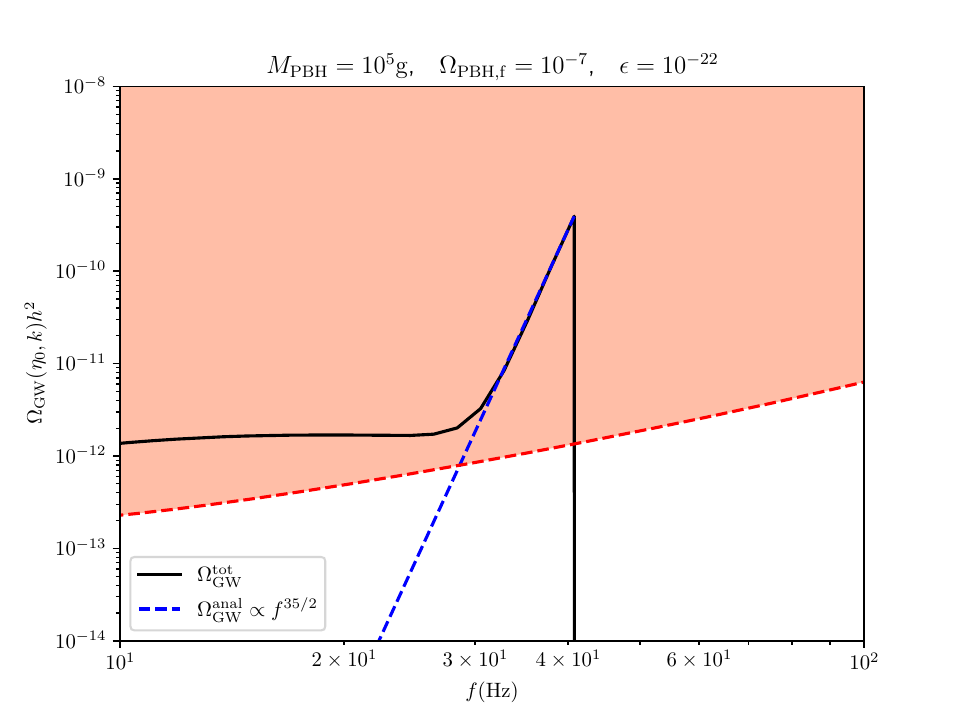}
\includegraphics[width = 0.49\textwidth]{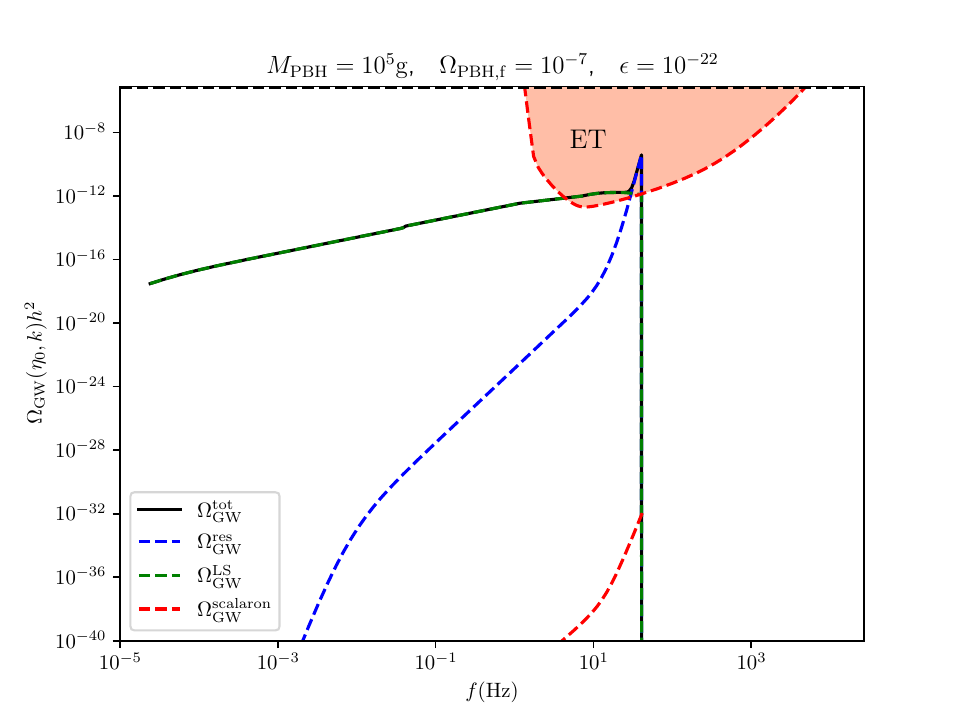}
\caption{{Left Panel: The scalar-induced gravitational wave signal in $R^{1+\epsilon}$ gravity for $M_\mathrm{PBH} = 10^5\mathrm{g}$, $\Omega_\mathrm{PBH,f} = 10^{-7}$ and $\epsilon = 10^{-22}$  (black solid curve) and the $f^{35/2}$ analytic approximation of the UV frequency region (dashed blue line). Right Panel: The scalar-induced gravitational wave signal in $R^{1+\epsilon}$ gravity for $M_\mathrm{PBH} = 10^5\mathrm{g}$, $\Omega_\mathrm{PBH,f} = 10^{-7}$ and $\epsilon = 10^{-22}$ (black solid curve) and the relevant dominant contributions to it, namely the resonant one (blue dashed line), the large $v$ (LV) one (green dashed line) and the scalaron one (red dashed line).}}
\label{fig:Omega_GW_ET_zoom}
\end{center}
\end{figure}

In the ultraviolet (UV) frequency however, one can check that within $R^{1+\epsilon}$ gravity, $\Omega_\mathrm{GW}\propto f^{35/2}$ around $k_\mathrm{c}$ [See the left panel of \Fig{fig:Omega_GW_ET_zoom}], in contrast with the respective $f^{11/3}$ GR frequency scaling~\cite{Domenech:2020ssp}. This highly peaky behaviour can be understood from the fact that at $k = k_\mathrm{c}$ all of the three GW dominant contributions (resonant, $\mathrm{LS}$ and scalaron ones) peak, mostly the resonant and large $v$ (LV) contributions [See the right panel of \Fig{fig:Omega_GW_ET_zoom}], making thus this $f(R)$ gravity GW signature quite distinctive compared to the respective GR one, where the SIGW signal features a smoother behaviour around its peak frequency at $k = k_\mathrm{UV}$.

At this point, one should make a comment as well regarding the nature of the scalaron contribution, which as we see from the right panel of \Fig{fig:Omega_GW_ET_zoom} is subdominant compared to the LV and resonant ones. The scalaron is actually not a transverse-traceless spin-$2$ field. Nevertheless, since it is a propagating degree of freedom obeying a wave equation it will contribute to the total GW energy density measured by a GW detector. This is why we ``add" in \Fig{fig:Omega_GW_ET_zoom} the GW spectrum of this scalar mode to the GW spectrum of the ``true" $(\times)$ and $(+)$ tensor polarisation GW modes. To the best of our knowledge, we cannot measure the scalar nature of an extra degree of freedom contributing to the GW energy density with the current GW observational apparatus, which only measures the gravitational strain. Ongoing works in this direction are still in progress~\cite{Baumgartner:2020oad,Wang:2021mou,Hu:2024toa}.

\begin{figure}
\begin{center}
\includegraphics[width = 0.6\textwidth]{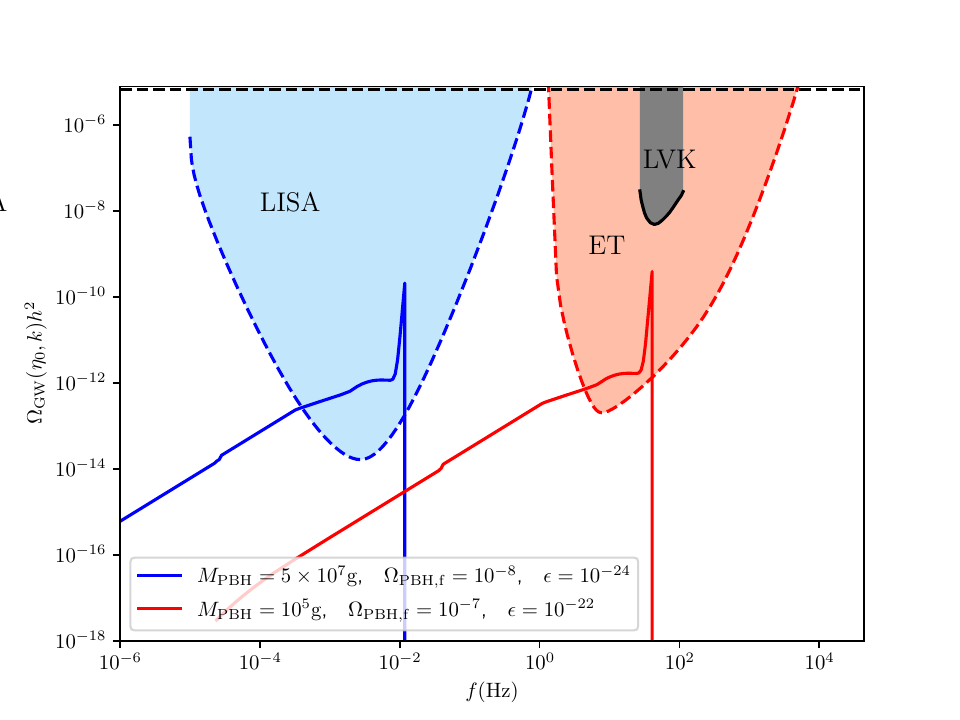}
\caption{{The scalar-induced gravitational wave signal in $R^{1+\epsilon}$ gravity for different values of the PBH mass $M_\mathrm{PBH}$, the initial PBH abundance $\Omega_\mathrm{PBH,f}$ and the $\epsilon$ parameter. The detection sensitivity curves of LIGO-VIRGO-KAGRA (LVK) \cite{KAGRA:2021kbb}, LISA \cite{LISA:2017pwj, Karnesis:2022vdp}, ET \cite{Maggiore:2019uih} and BBO \cite{Harry:2006fi} are also superimposed for comparison.}}
\label{fig:Omega_GW_FR}
\end{center}
\end{figure}

Followingly, in \Fig{fig:Omega_GW_FR}, we show the SIGW signal in $R^{1+\epsilon}$ gravity for different values of the PBH mass $M_\mathrm{PBH}$, the initial PBH abundance $\Omega_\mathrm{PBH,f}$ and the $\epsilon$ parameter, superimposing as well the detection sensitivity curves of LIGO-VIRGO-KAGRA (LVK) \cite{KAGRA:2021kbb}, LISA \cite{LISA:2017pwj, Karnesis:2022vdp}, ET \cite{Maggiore:2019uih} and BBO \cite{Harry:2006fi} GW observatories. For all the figures we have set $M_\mathrm{s}=10^{-5}\Mp$. We checked that by varying $M_\mathrm{s}$ there is no difference at the level of the spectral shape of the signal and the GW amplitude. Interestingly enough, for reasonable choices of our free parameters one can get a GW signal within the detection bands of GW experiments, rendering thus this signal potentially detectable in the future.

\begin{figure}
\begin{center}
\includegraphics[width = 0.8\textwidth]{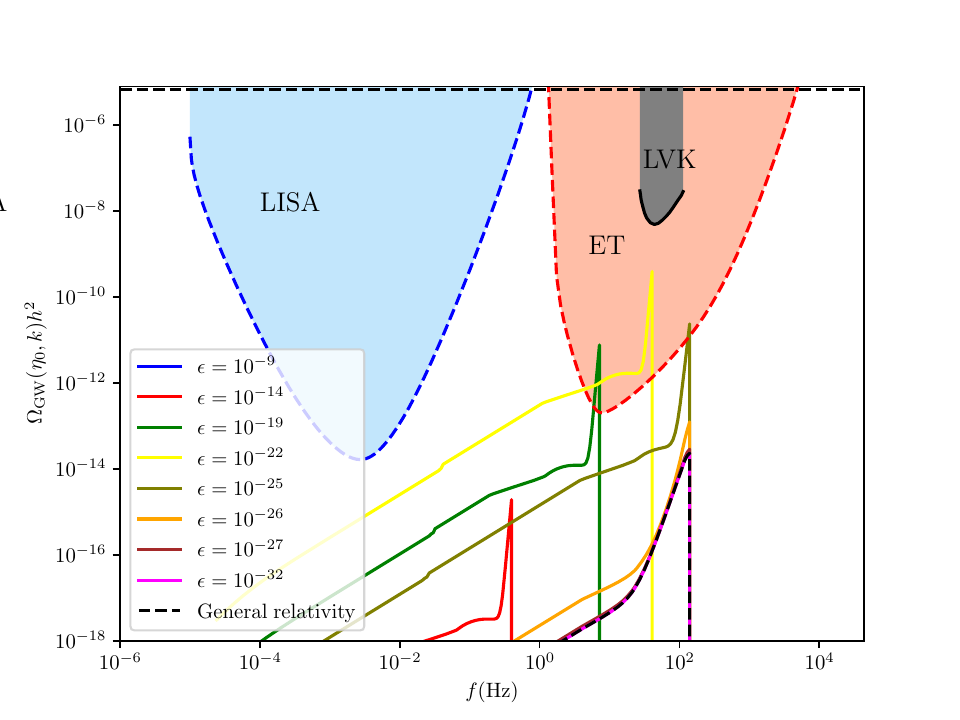}
\caption{{The scalar-induced gravitational wave signal in $R^{1+\epsilon}$ gravity for $M_\mathrm{PBH} = 10^5\mathrm{g}$ and $\Omega_\mathrm{PBH,f} = 10^{-7}$ for different values of the parameter $\epsilon$ within the frequency range probed by the Einstein Telescope~\cite{Maggiore:2019uih}}.}
\label{fig:Omega_GW_epsilon_comparison}
\end{center}
\end{figure}

Finally, let us see how the induced GW signal varies with the parameter $\epsilon$. For concreteness, we will work here with $M_\mathrm{PBH} = 10^5\mathrm{g}$ and $\Omega_\mathrm{PBH,f}=10^{-7}$ as we did also in the previous figures, focusing in the ET frequency range. In \Fig{fig:Omega_GW_epsilon_comparison}, we see how our GW signal varies with the parameter $\epsilon$. In particular, as one decreases the value of $\epsilon$ the GW amplitude is increasing taking its maximum value at $\epsilon\sim 10^{-22}$. Then, as we continue decrease $\epsilon$ we converge gradually to GR with the perfect convergence happening for $\epsilon\lesssim 10^{-25}$.

\begin{figure}
\begin{center}
\includegraphics[width = 0.49\textwidth]{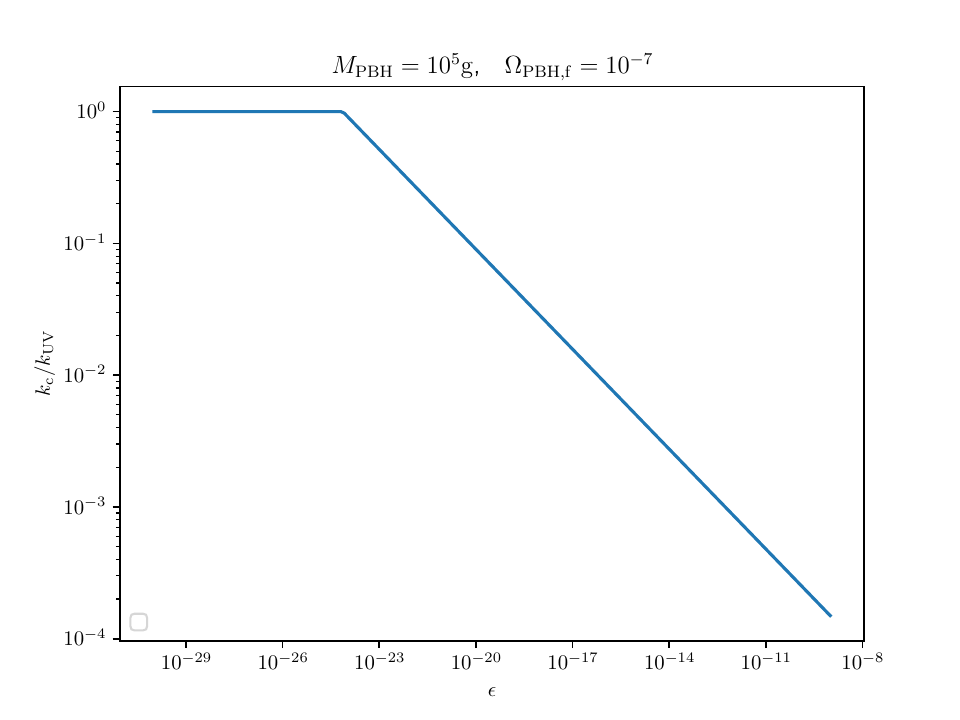}
\includegraphics[width = 0.49\textwidth]{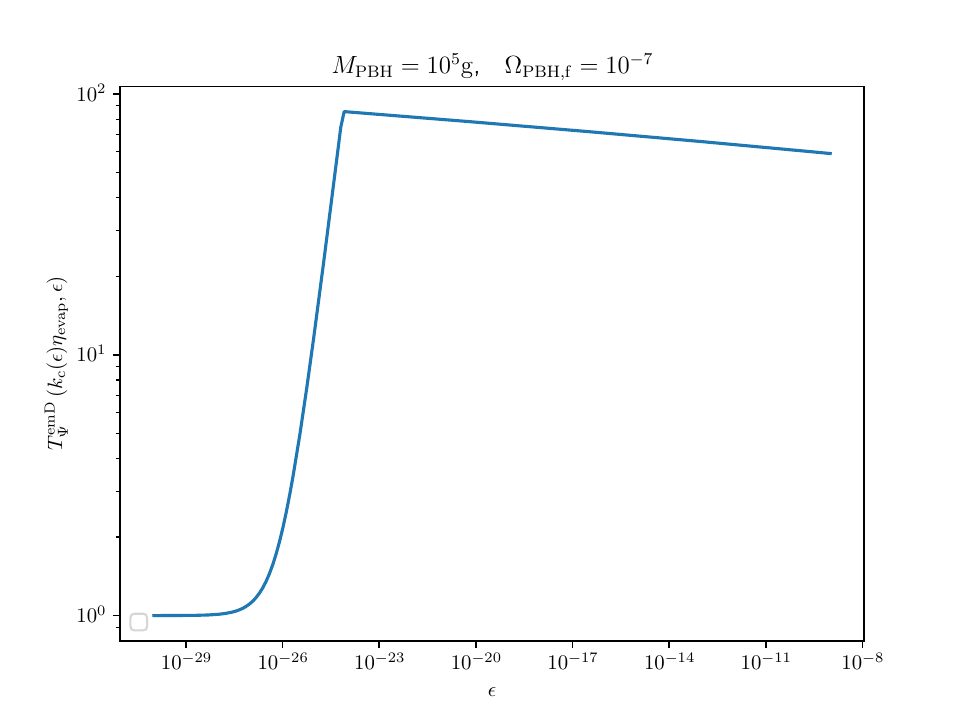}
\caption{{Left Panel: The ratio $k_\mathrm{c}/k_\mathrm{UV}$ for $M_\mathrm{PBH} = 10^5\mathrm{g}$, $\Omega_\mathrm{PBH,f} = 10^{-7}$ as a function of $\epsilon$. Right Panel: The PBH-eMD transfer function at the end of the PBH-eMD era, $T^\mathrm{eMD}\left(k_\mathrm{c}(\epsilon)\eta_\mathrm{evap}, \epsilon\right)$ at $k=k_\mathrm{c}(\epsilon)$ as a function of $\epsilon$.}}
\label{fig:k_c_T_Psi_kc}
\end{center}
\end{figure}

This non-monotonic convergence to the GR GW spectrum as $\epsilon\rightarrow 0$ can be justified if we see more carefully the places in the GW computation where one observes an $\epsilon$ dependence. In particular, there are two contributions to the GW spectrum where we see an $\epsilon$ dependence competing each other. Firstly, $\epsilon$ enters in the determination of $k_\mathrm{c}$, i.e. the non-linear scale below which the power spectrum $\mathcal{P}_\Psi$ becomes greater than unity. As one decreases $\epsilon$, $\mathcal{P}_\Psi$ becomes less and less non-linear and as a consequence smaller scales (equivalently higher values of $k_\mathrm{c}$) contribute to the double integrals \Eq{eq:Omega_GW_res_full} and \Eq{eq:Omega_GW_LV_full} leading to an increase of the GW amplitude. On the other hand, $\epsilon$ enters as well in the determination of $T_{\Psi,\mathrm{eMD}}(k\eta,\epsilon)$ which for $k=k_\mathrm{c}(\epsilon)$ (the dominant contribution), gets an $\epsilon$ dependence in both of its arguments competing each other. Namely, if one decreases $\epsilon$, $k_\mathrm{c}$ increases [See the left panel of \Fig{fig:k_c_T_Psi_kc}] leading to an increase of $T_{\Psi,\mathrm{eMD}}$. However, a decrease of the value of $\epsilon$ in the second argument of $T_{\Psi,\mathrm{eMD}}$ will lead to a decrease of $T_{\Psi,\mathrm{eMD}}$ [See \Fig{fig:T_Psi_eMD_vs_epsilon}].

Ultimately, one can infer then that as we decrease $\epsilon$ the $k_\mathrm{c}$ contribution always wins and the GW amplitude increases, with $T_{\Psi,\mathrm{eMD}}$ being of the order of $60-80$ for a large range of $\epsilon$ up to $10^{-25}$ [See the right panel of \Fig{fig:k_c_T_Psi_kc}]. This happens because $T_{\Psi,\mathrm{eMD}}(k_\mathrm{c}(\epsilon)\eta_\mathrm{evap},\epsilon)$ depends on $\epsilon$ in both of each arguments, $k_\mathrm{c}(\epsilon)\eta_\mathrm{evap}$ and $\epsilon$, and for large values of $\epsilon>10^{-25}$, the $k_\mathrm{c}(\epsilon)\eta_\mathrm{evap}$ term dominates. If one continues now to decrease $\epsilon$ they will see that for $\epsilon<10^{-25}$, $T_{\Psi,\mathrm{eMD}}$ decreases abruptly to unity [See the right panel of \Fig{fig:k_c_T_Psi_kc}], leading to a perfect convergence to GR for $\epsilon \lesssim 10^{-25}$ when $k_\mathrm{c}\rightarrow k_\mathrm{UV}$ [See also the left panel of \Fig{fig:k_c_T_Psi_kc}].

\section{Discussion and Conclusions}\label{sec:conclusions}

In this work, we studied the GW signal induced by PBH isocurvature energy density perturbations within the context of $f(R)$ gravity, following the recent developments in the field of scalar-induced GWs in modified gravity~\cite{Zhou:2024doz,Kugarajh:2025rbt,Lopez:2025gfu}. In particular, we extended significantly a previous work~\cite{Papanikolaou:2021uhe} on the topic accounting for the full modified source of SIGWs being characterized by an extra geometry-induced anisotropic stress. We also thoroughly studied the evolution of scalar perturbations during an eMD era driven by ultra-light PBHs in the framework of  $f(R)$ gravity, studying carefully super and sub-horizon scales. We also accounted, for the first time, for the effect of the scalaron massive mode on the first-order GW background.

We performed our analysis by focusing on two particular $f(R)$ gravity models, namely  the $R^2$ gravity, giving rise to a viable inflation scenario and being in very good agreement with Planck observations as well as the $R^{1+\epsilon}$ gravity with $\epsilon\ll 1$, naturally accounting for regularization and renormalization issues of GR up to one-loop level. Interestingly, $R^2$ gravity was found to have negligible effects at the level of both the scalar and the tensor perturbations compared to GR, something which is in agreement with the findings in~\cite{Papanikolaou:2021uhe}. However, $R^{1+\epsilon}$ gravity was found to have a totally different behaviour exhibiting, in particular, an exponential growth of sub-horizon scalar perturbations during the PBH-driven eMD era making us to impose a non-linear cut-off scale $k^{-1}_\mathrm{c}$ below which perturbation theory breaks down, larger compared to the GR cut-off scale.

This unique feature of $R^{1+\epsilon}$ gravity led inevitably to an enhanced induced GW signal on smaller frequencies compared to the frequencies where the respective GR signal peaks. %as well as to an enhanced GW amplitude even for very small initial PBH abundances. 
Furthermore, it was found that accounting for all the dominant contributions to the GW signal, the induced GW spectrum shares a universal linear frequency scaling on small frequencies away from its peak, namely $\Omega_\mathrm{GW}\propto f$ for $f\ll f_\mathrm{peak} = k_\mathrm{c}/(2\pi)$, as in the case of GR, but on large frequencies close to the non-linear cut-off scale $k_\mathrm{c}$, scales as $f^{35/2}$, in contrast with the GR $f^{11/3}$ frequncy scaling.

Our work brought together the fields of PBHs, GWs and modified gravity opening new prespectives in searching for distinctive signatures of modified gravity theories through the portals of PBHs and GWs that can make them distinguishable compared to GR. Notably, the exponential growth of scalar perturbations during a MD era in $R^{1+\epsilon}$ gravity is just a characteristic example of a distinctive signature for $f(R)$ gravity, which can have a significant impact for early structure formation in the early Universe~\cite{Jedamzik:2010dq,Barenboim:2013gya,Eggemeier:2020zeg} and the associated to it GW phenomenology~\cite{Dalianis:2020gup,Fernandez:2023ddy,Dalianis:2024kjr}.

\section*{Acknowledgments}
The authors acknowledge stimulating discussions with Charalampos Tzerefos as well as the support of  Istituto Nazionale di Fisica Nucleare (INFN), Sez.  di Napoli,  {\it Iniziative Specifiche} QGSKY and MOONLIGHT2. S.\,C. is grateful to the {\it Gruppo Nazionale di Fisica Matematica} (GNFM) of {\it Istituto Nazionale di Alta Matematica} (INDAM) for the support. 
This paper is supported by the COST Action CA21136 {\it Addressing observational tensions in cosmology with systematics  and fundamental physics} (CosmoVerse) supported by COST (European Cooperation in Science and Technology). TP receives as well financial support by the funding program 
“MEDICUS”  of the University of Patras.

\newpage
\appendix

\section{Evolution of $\delta_\mathrm{PBH}$ in $f(R)$ gravity}\label{app:delta_PBH}
In this Appendix we study the evolution of $\delta_\mathrm{PBH}$ within  $f(R)$ gravity, we consider here, namely within $R^2$ and $R^{1+\epsilon}$ gravity. Following the relevant analysis in~\cite{Tsujikawa:2007gd,Papanikolaou:2021uhe} one can show that the respective growth equation for $\delta_\mathrm{PBH}$ in the sub-horizon regime, namely the M\'eszaros equation, will read as
\bea\label{eq:Meszaros in f(R):subhorizon}
\frac{\dd^2 \delta_\mathrm{PBH}}{\dd s^2}+\frac{2+3s}{2s(s+1)}\frac{\dd \delta_\mathrm{PBH}}{\dd s}-\frac{3}{2s (s+1)} \frac{1}{F} \frac{1 +4\frac{k^2}{a^2}\frac{F_\mathrm{,R}}{F}}{1+3\frac{k^2}{a^2}\frac{F_\mathrm{,R}}{F}}\delta_\mathrm{PBH}=0\,.
\eea

\begin{figure}[ht]
\centering
\includegraphics[width=0.496\textwidth]{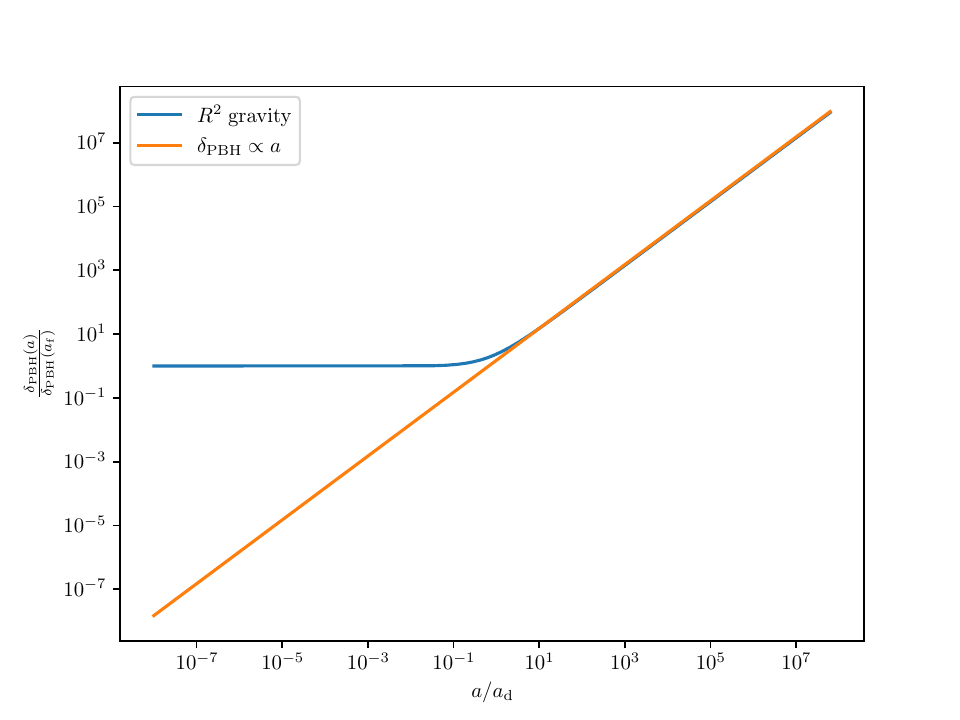}
\includegraphics[width=0.496\textwidth]{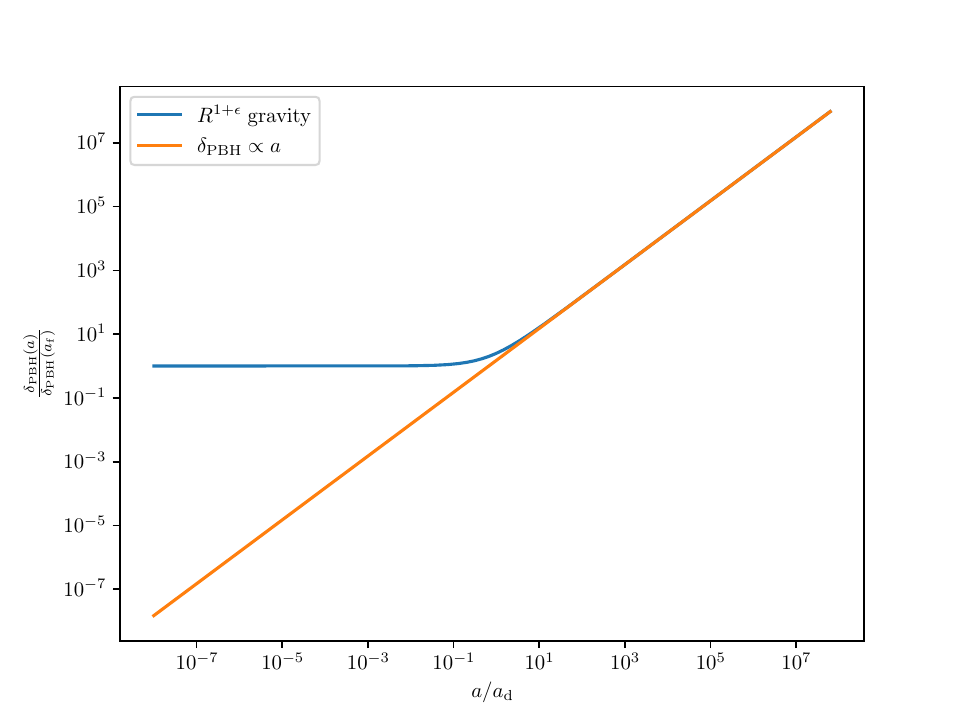}
\caption{\it{\underline{Left Panel}: The evolution of $\delta_\mathrm{PBH}(a)/\delta_\mathrm{PBH}(a_\mathrm{f})$ as a function of $a/a_\mathrm{d}$ in $R^2$ gravity for $M_\mathrm{PBH} = 10^4\mathrm{g}$, $\Omega_\mathrm{PBH,f}=10^{-7}$, $M_\mathrm{s}=10^{-5}\Mp$ and $k=k_\mathrm{d}$. \underline{Right Panel}: The evolution of $\delta_\mathrm{PBH}(a)/\delta_\mathrm{PBH}(a_\mathrm{f})$ as a function of $a/a_\mathrm{d}$ in $R^{1+\epsilon}$ gravity for $M_\mathrm{PBH} = 10^4\mathrm{g}$, $\Omega_\mathrm{PBH,f}=10^{-7}$, $M_\mathrm{s}=10^{-5}\Mp$ and $k=k_\mathrm{d}$.}}
\label{fig:delta_PBH}
\end{figure}

We show below in \Fig{fig:delta_PBH} with the blue curves the evolution of $\delta_\mathrm{PBH}(a)/\delta_\mathrm{PBH}(a_\mathrm{f})$ as a function of $a/a_\mathrm{d}$ in both $R^2$ and $R^{1+\epsilon}$ gravity for $M_\mathrm{PBH} = 10^4\mathrm{g}$, $\Omega_\mathrm{PBH,f}=10^{-7}$, $M_\mathrm{s}=10^{-5}\Mp$ and $k=k_\mathrm{d}$ superimposing as well in the orange curves the GR evolution of $\delta_\mathrm{PBH}$ growing linearly with the scale factor. As one may infer, $\delta_\mathrm{PBH}$ in both $R^2$ and $R^{1+\epsilon}$ gravity grows linearly with the scale factor following the GR behaviour. This behaviour does not change as checked numerically if we vary the parameters $M_\mathrm{PBH}$, $\Omega_\mathrm{PBH,f}$, $M_\mathrm{s}$ and the comoving number $k$, justifying thus the linear growth of $\delta_\mathrm{PBH}$ in \Sec{sec:Psi_Phi_PBH_f_R_gravity} when extracting the power spectra $\mathcal{P}_\Psi(k)$ and $\mathcal{P}_\Phi(k)$.

\newpage
\section{Continuity of $\Psi$ and $\Psi^{\prime}$ in the Newtonian gauge}\label{app:Synchronous_gauge}
The energy transfer due to PBH evaporation is in general non-local since in depends on the spatial coordinates through the PBH 4-velocity~\cite{Domenech:2020ssp}. It becomes local only within the synchronous comoving gauge where the scalarly perturbed metric can be written as 
\beq\label{eq:perturbed_metric_syncronous}
\mathrm{d}s^2=a^2(\tau)\left[-\mathrm{d}\eta^2+\left(\delta_{ij}-2\Psi_\mathrm{s}\delta_{ij}+2\partial_i\partial_j E_\mathrm{s}\right)\mathrm{d}x^i\mathrm{d}x^j\right].
\eeq

Within this work we performed the calculation of the induced GW signal within the Newtonian gauge since the analysis simplifies a lot within this gauge. We imposed thus by hand the continuity of $\Psi$ and $\Psi'$ in the Newtonian gauge. However, one should check that the continuity of the Bardeen gravitational potentials and their derivatives in the physical synchronous comoving gauge, where PBH evaporation is purely local, is equivalent with the continuity of $\Psi$ and $\Psi'$ in the Newtonian gauge as well.

Focusing firstly in GR, one can show that after a gauge transformation, $\Psi$ in the Newtonian gauge is related to $\Psi$ in the synchronous comoving gauge as follows:
\beq\label{eq:Psi_N_vs_Psi_s}
\Psi_\mathrm{N} = \Psi_\mathrm{s} - \mathcal{H}E'_\mathrm{s}
\eeq
where the subscripts $\mathrm{s}$ and $\mathrm{N}$ stand for the synchronous comoving and the Newtonian gauges respectively.
In particular, one has that the continuity of the metric components and their time derivatives in the synchronous comoving gauge will imply by definition the
continuity of $\Psi_\mathrm{s}$ and $E_\mathrm{s}$ and their derivatives. However, from \Eq{eq:Psi_N_vs_Psi_s}, one gets the continuity of $\Psi_\mathrm{N}$ from the continuity of $\Psi_\mathrm{s}$ and $E^\prime_\mathrm{s}$. In order thus to get the continuity for
$\Psi'_\mathrm{N}$ we need to have the continuity of $E^{''}_\mathrm{s}$. This is the case for GR if one considers the traceless-$ij$ component of the perturbed field equations in the synchronous comoving gauge giving us that
\beq
E^{\prime\prime}_\mathrm{s} + 2\mathcal{H}E'_\mathrm{s} = \Psi_\mathrm{s}
\eeq
from which we see that the continuity of $\Psi_\mathrm{s}$ and $E'_\mathrm{s}$ entails the continuity  $E^{\prime\prime}_\mathrm{s}$.

Going to the case of $f(R)$ gravity we will consider as an illustrative example the case of $R^2$ gravity where the calculation is simpler but our findings will be independent for any power-law $f(R)$ gravity theory. In particular, following the same reasoning as before one obtains from the traceless-$ij$ component of the perturbed field equations in the synchronous comoving gauge that
\beq
\begin{split}
& -2\left[(\mathcal{H}^2 + 2\mathcal{H}') - \frac{6\mathcal{H}^4}{M^2_\mathrm{s}a^2}-\frac{2\mathcal{H}^{'2}}{M^2_\mathrm{s}a^2} -\frac{4\mathcal{H}^{'''}}{M^2_\mathrm{s}a^2} + \frac{24\mathcal{H}^2\mathcal{H}'}{M^2_\mathrm{s}a^2}+ \frac{4\mathcal{H}\mathcal{H}^{''}}{M^2_\mathrm{s}a^2}\right]E_\mathrm{s} \\ & + \left[2\mathcal{H} + \frac{2\mathcal{H}^{''}}{M^2_\mathrm{s}a^2} + \frac{4\mathcal{H}\mathcal{H}'}{M^2_\mathrm{s}a^2} - \frac{2\mathcal{H}\nabla^2}{M^2_\mathrm{s}a^2}\right]E^{\prime}_\mathrm{s} + \left[1 + \frac{2\mathcal{H}^\prime + 2\mathcal{H}^2- \frac{2\nabla^2}{3}}{M^2_\mathrm{s}a^2}\right]E^{\prime\prime}_\mathrm{s}
\\ & + \left[1 + \frac{2\mathcal{H}^\prime}{M^2_\mathrm{s}a^2} + \frac{2\mathcal{H}^2}{M^2_\mathrm{s}a^2}  - \frac{4\nabla^2}{3M^2_\mathrm{s}a^2}   \right]\Psi_\mathrm{s}  + \frac{6\mathcal{H}}{M^2_\mathrm{s}a^2}\Psi^\prime_\mathrm{s} + \frac{2}{M^2_\mathrm{s}a^2}\Psi^{\prime\prime}_\mathrm{s} = 0,
\end{split}
\eeq
from which we see that the continuity of $\Psi_\mathrm{s}$, $\Psi^\prime_\mathrm{s}$, $E_\mathrm{s}$ and $E^\prime_\mathrm{s}$ imply the continuity of $\Psi^{\prime\prime}_\mathrm{s}$ and $E^{\prime\prime}_\mathrm{s}$. One thus gets again that in $f(R)$ gravity the continuity of $\Psi$ and $\Psi^\prime$ in the physical synchronous comoving gauge implies the continuity of $\Psi$ and $\Psi^\prime$ in the Newtonian gauge as well.

\bibliographystyle{JHEP} 
\bibliography{PBH}
\end{document}